\documentclass[a4paper,11pt]{article}
\pdfoutput=1

\usepackage{jheppub}
\usepackage{url}
\usepackage[nameinlink]{cleveref}
\usepackage{orcidlink}
\usepackage[T1]{fontenc}
\usepackage{dsfont}
\usepackage{subcaption}
\usepackage{slashed}
\usepackage{pdflscape}
\usepackage{placeins}
\usepackage{bbold}
\usepackage{afterpage}

\usepackage{enumitem}
\setitemize{fullwidth}

\hypersetup{colorlinks=true}

\newcommand{\nn}{\nonumber}
\NewDocumentCommand \dd { o m }{
  \IfNoValueTF {#1}
    {{\mathrm{d} #2\;}}
    {{\mathrm{d}^{#1} #2\;}}
}
\newcommand{\STr}{\mathrm{STr}\,}
\newcommand{\barsigma}{\bar{\sigma}}
\newcommand{\hc}{\mathrm{h.c.}}
\newcommand{\levicivita}{\varepsilon}
\newcommand{\ii}{\mathrm{i}}
\newcommand{\cqq}{c_{qq}}
\newcommand{\cqu}{c_{qu}}
\newcommand{\cqd}{c_{qd}}
\newcommand{\cuu}{c_{uu}}
\newcommand{\cud}{c_{ud}}
\newcommand{\cdd}{c_{dd}}
\newcommand{\cquqd}{c_{quqd}}
\newcommand{\cRe}{c_{quqd}^{+}}
\newcommand{\cIm}{c_{quqd}^{-}}
\newcommand{\CRe}{C_{quqd}^{+}}
\newcommand{\CIm}{C_{quqd}^{-}}

\newcommand{\set}[1]{\left\{\,#1\,\right\}}

\newcommand{\Dleftright}{\!\!\stackrel{\leftrightarrow}
{D}_\mu\!\!}
\newcommand{\DIleftright}{\!\!\stackrel{\leftrightarrow}{D}\!\!{}^{\,I}_\mu}
\newcommand{\dotalpha}{{\dot{\alpha}}}
\newcommand{\dotbeta}{{\dot{\beta}}}
\newcommand{\dotgamma}{{\dot{\gamma}}}

\title{The fermion sector of the SMEFT from asymptotically safe gravity
}

\author[a]{Astrid Eichhorn \orcidlink{0000-0003-4458-1495},} 
\author[a]{Moritz Gessner \orcidlink{0009-0001-2342-9606}}
\author[b]{and Shouryya Ray \orcidlink{0000-0003-4754-0955}}

\affiliation[a]{Institut f\"ur Theoretische Physik, Universit\"at Heidelberg, Philosophenweg 12 \& 16, 69120 Heidelberg, Germany}
\affiliation[b]{N\'att\'uruv\'isindadeildin, Fr\o{}\dh{}skaparsetur F\o{}roya, Fr\ae{}lsi\dh{} 20, 100 T\'orshavn, Faroe Islands}

\emailAdd{eichhorn@thphys.uni-heidelberg.de}
\emailAdd{gessner@thphys.uni-heidelberg.de}
\emailAdd{shouryyar@setur.fo}

\abstract{Quantum gravity impacts Standard Model fields through effective interactions generated by renormalization.
These appear as Standard Model Effective Field Theory (SMEFT) operators, where the effects of new physics are parametrized by the values of the higher-order SMEFT coefficients.
As a step towards predicting these SMEFT coefficients from the asymptotically safe Standard Model with quantum gravity, we investigate the flow of a representative set of four-fermion operators under gravitational fluctuations. 
We strengthen the evidence for the near-perturbative nature of asymptotic safety by finding that these dimension-six-interactions remain irrelevant, resulting in the prediction of specific values for the dimensionless ratios of SMEFT coefficients in the IR.
We also find a mechanism that can make a specific subset of these operators relevant.
This subset is determined by symmetry considerations.
On this basis, we provide  a categorization of dimension-six-SMEFT interactions into two categories.
Interactions in the first category  are predicted to be non-zero, but Planck-scale suppressed.
Interactions in the second category are predicted to be zero at small gravitational coupling, but may become free parameters of the theory at larger gravitational couplings.
}

\begin{document}
\maketitle

\section{Introduction}

The Standard Model (SM) of particle physics is an effective field theory in need of an ultraviolet (UV) completion.
This UV completion contains new degrees of freedom, that become important beyond a scale of new physics $\Lambda_{\rm UV}$.
Before the discovery of the Higgs boson at the LHC \cite{CMS:2012qbp, ATLAS:2012yve}, it was unclear whether this scale would be as low as a few TeV.\footnote{Here and henceforth, when we refer to a ``scale'', we exclusively mean mass, energy, or momentum scales.
Thus, the UV corresponds to ``high'' or ``large'' values of this scale, whereas the IR corresponds to ``low'' or ``small'' scales.
Note that authors who use ``scale'' as in length scales use the same modifiers to mean the exact opposite with respect to the UV and the IR.} At high values of the Higgs mass, a Landau pole in the Higgs quartic coupling signals the need of new physics. 
The scale of the Landau pole bounds the scale of new physics from above.
At a Higgs mass of 125 GeV, this Landau pole is avoided.
Instead, the SM remains perturbative and self-consistent all the way to the Planck scale.\footnote{A metastable electroweak vacuum that may be realized, depending on the exact value of the top quark mass \cite{Markkanen:2018pdo}, but has a lifetime that exceeds the age of our universe, not threatening the phenomenological viability of the SM.} In this scenario, the first indication that new physics is needed arises at a transplanckian scale, where the U(1) hypercharge coupling hits a Landau pole.
Thus, a key takeaway from the LHC observations is that there is no definite need for new physics before below the Planck scale, only above.
Therefore, a particularly simple possibility is that the SM together with quantum gravity becomes UV complete.\footnote{Dark matter may be extremely weakly coupled to the SM, such that it does not interfere with this conclusion.} 

A strong candidate for such a UV completion is asymptotically safe quantum gravity.
Originally proposed by Weinberg \cite{Weinberg:1980gg}, and technically accessible in four dimensions since a seminal paper by Reuter \cite{Reuter:1996cp}, the field of asymptotically safe quantum gravity has been undergoing a significant development over the last decade.
By now, asymptotic safety is regarded as established for practical purposes in four-dimensional, Euclidean, pure gravity, see \cite{Niedermaier:2006wt,Reuter:2012id,Eichhorn:2018yfc,Pereira:2019dbn,Pawlowski:2020qer,Eichhorn:2022jqj,Eichhorn:2022gku,Eichhorn:2022bgu,Wetterich:2022ncl,Knorr:2022dsx,Morris:2022btf,Martini:2022sll,Pawlowski:2023gym,Saueressig:2023irs,Platania:2023srt,Bonanno:2024xne,Eichhorn:2023xee} for reviews, \cite{Percacci:2017fkn,Reuter:2019byg}
for books and \cite{Eichhorn:2020mte,Reichert:2020mja,Basile:2024oms} for lecture notes.
The most up-to-date review is \cite{Eichhorn:2026uqj}.
The study of Lorentzian signature, long an important open issue, has by now yielded first positive results \cite{Fehre:2021eob,DAngelo:2022vsh,DAngelo:2023tis,Pawlowski:2025etp,Kher:2025rve,Assant:2026dca}, supported by results in Euclidean signature that admit an analytical continuation \cite{Manrique:2011jc,Rechenberger:2012dt,Biemans:2016rvp,Houthoff:2017oam,Biemans:2017zca,Saueressig:2023tfy,Korver:2024sam,Saueressig:2025ypi,Bonanno:2021squ,Pastor-Gutierrez:2024sbt,Chiesa:2026tlz,Knorr:2026jcg}.
Most importantly for the purposes of this paper, an increasing body of literature is devoted to the interplay of asymptotically safe quantum gravity with matter, as reviewed in \cite{Eichhorn:2022jqj,Eichhorn:2022gku,Eichhorn:2026uqj}.
Key results include the observation that asymptotic safety in gravity is left intact under the impact of matter fields \cite{Dona:2013qba}, see also \cite{Meibohm:2015twa, Biemans:2017zca,Alkofer:2018fxj, Wetterich:2019zdo, Pastor-Gutierrez:2022nki}; and that asymptotic safety has predictive power for Standard Model couplings, including the Higgs quartic coupling \cite{Shaposhnikov:2009pv}, the Abelian gauge coupling \cite{Harst:2011zx, Eichhorn:2017lry} and a subset of Yukawa couplings and mixing matrix entries \cite{Eichhorn:2017ylw,Eichhorn:2018whv,Eichhorn:2025sux}.

Against this background, it is vital to understand the structure of the asymptotically safe Standard Model, i.e., the SM with gravity, in more detail.
In particular, after the possible effects on the leading order, dimension-four-interactions \cite{Shaposhnikov:2009pv, Harst:2011zx, Eichhorn:2017lry}, see also \cite{Daum:2009dn,Zanusso:2009bs,Daum:2010bc,Vacca:2010mj,Folkerts:2011jz,Eichhorn:2016esv,Eichhorn:2017als,Christiansen:2017gtg,Eichhorn:2017eht,Eichhorn:2017ylw,Pawlowski:2018ixd,Eichhorn:2020sbo,Alkofer:2020vtb,Kowalska:2022ypk,Eichhorn:2022vgp,deBrito:2025nog,Riabokon:2025ozw,Eichhorn:2025sux}, it is the logical next step to study higher-order interactions.
The Standard Model Effective Field Theory (SMEFT) \cite{Buchmuller:1985jz,Grzadkowski:2010es,Brivio:2017vri,Isidori:2023pyp,Falkowski:2023hsg} is a framework taylor-made for this purpose.
The SMEFT is based on the fields and symmetries of the SM, and extends the interactions of those fields to higher order in the mass dimension of the operators.
At each order in the mass dimension, the SMEFT consists of all possible (quasi-local) operators compatible with the symmetries of the SM, parametrized by Wilson coefficients of unknown value.
Wilson coefficients become non-zero, when UV degrees of freedom beyond a scale $\Lambda_{\text{UV}}$ are integrated out.
Their value depends on the UV completion.
Thus, experimental constraints on the UV completion can be derived by measuring (or constraining) Wilson coefficients, even if the scale of the UV completion is significantly higher than experimental scales.

In this spirit, robust predictions for SMEFT coefficients are a first step towards experimental access to quantum gravity, using only experimental data at scales far below the Planck scale.
Current experimental bounds on SMEFT coefficients are compatible with zero \cite{Ellis:2021kzk, Isidori:2023pyp}, but only upper bounds exist \cite{Cirigliano:2012ab,deBlas:2013qqa,Falkowski:2015jaa,Greljo:2017vvb,Falkowski:2017pss,daSilvaAlmeida:2018iqo,vanBeek:2019evb,Ellis:2020unq,Ethier:2021bye,Greljo:2021kvv,Breso-Pla:2023tnz,Kassabov:2023hbm,Coloma:2024ict,Celada:2024oax,Celada:2024mcf}, see also \cite{Aebischer:2025qhh} and references therein.

In the scenario of a ``desert'' between the electroweak scale and the Planck scale, the leading order contribution to SMEFT coefficients comes directly from quantum gravity.\footnote{Even with the inclusion of beyond SM physics, e.g., Grand Unified Theories, not all SMEFT operators are necessarily generated by a specific UV completion.
Symmetry constraints might prevent certain Wilson coefficients from being non-zero for a given theory.
In this case, the leading contribution to a deviation from zero might still be given by quantum gravity.}
Without further concrete experimental hints for physics beyond the SM, it is therefore incumbent upon us to study
the values of SMEFT Wilson coefficients in the existing frameworks for quantum gravity.
Based on standard effective-field-theory (EFT) reasoning, one expects that the ``desert'' scenario results in unmeasurably tiny, Planck-scale suppressed, Wilson coefficients. 
In addition, this implies that the coefficients of dimension-eight-operators are very strongly suppressed compared to the coefficients of dimension-six-operators and so on.
It is clearly critical to check this expected EFT ordering with concrete calculations within a quantum-gravity framework. Thus, the first question that we will address is the following: in a given quantum-gravity framework, is the EFT ordering principle for SMEFT coefficients respected? In other words, do SMEFT coefficients take their ``natural'' values expected within a Planckian UV completion?

Asymptotically safe quantum gravity is a candidate theory of quantum gravity with a technically and conceptually clear access to IR observables. It is a quantum field theory for the metric and the coupling to the matter fields of the SM is straightforward.
This theory is therefore in an optimal position to study SMEFT operators. 
Thus, considerations for particular SMEFT operators in asymptotic safety have been investigated previously in \cite{Eichhorn:2023jyr, Brenner:2024bps}.
There, it was found that the generic expectation, that the SMEFT coefficients are Planck-scale suppressed, cannot be circumvented, unless asymptotic safety provides a truly non-perturbative UV completion and results in non-perturbative effects, encoded in non-perturbative RG fixed points, even below the Planck scale. Here, we aim at extending such a conclusion to a much larger set of operators and making it more robust.

In this paper, we therefore take a step towards a general classification of
SMEFT operators with respect to their behavior under a UV completion by asymptotically safe quantum gravity.
This classification is based on \cite{Eichhorn:2017eht}.
To arrive at such a general classification, we focus on four-fermion interactions, which form the largest group of SMEFT operators at dimension six. We treat them both as a particularly important part of the SMEFT, that is vital to understand for its own sake, as well as a paradigmatic example of SMEFT operators which exemplify the response of the SMEFT coefficients to quantum gravity. Thus, our discussion of four-fermion interactions generalizes also to other sectors of the SMEFT.

Besides their role as a paradigm-setting example, four-fermion interactions are of particular interest at the phenomenological level. They constitute a diagnostic tool to track
the formation of bound states and spontaneous breaking of symmetries related to fermionic mass-generation. In general, large (or diverging) four-fermion couplings signal the condensation of a fermion bilinear, typically linked to the breaking of symmetries and formation of bound states. If such a scenario were realized in quantum gravity, the resulting bound states would be expected to be of Planckian mass. This would clearly be in stark contrast to the SM, where bound-state formation only sets in close to the QCD scale.
Therefore, four-fermion interactions enable us to understand whether the existence of fermions which are light compared to the Planck mass is actually compatible with asymptotic safety. 
In this spirit, references \cite{Eichhorn:2011pc, Meibohm:2016mkp,Eichhorn:2017eht, deBrito:2020dta, deBrito:2023kow} have previously studied four-fermion-interactions. However, they have exclusively focused on Dirac fermions, as opposed to Weyl fermions. 
As a result, the symmetry structure of the matter sector of the SM has not been properly accounted for. 
It is therefore vital to extend these studies to Weyl fermions, which constitute the fundamental fermion fields in the SM.

In summary, the questions that we will address in this paper are:
\begin{enumerate}
\item[1)] Does asymptotically safe quantum gravity follow the EFT ordering principle for SMEFT coefficients, with all SMEFT coefficients suppressed by powers of the Planck scale dictated by the mass dimension of the corresponding operator?
\item[2)] Do conclusions on the lightness of fermions, drawn in previous work on the basis of a simpler fermionic field content, persist, once Weyl fermions with distinct  charges for left- and right-handed fermions are used in the construction of four-fermion interactions?
\end{enumerate}

This paper is structured as follows: 
In Sec.~\ref{sec:prerequisites} we introduce the Standard Model effective field theory and review key aspects of asymptotically safe quantum gravity.
Sec.~\ref{sec:action} specifies our toy model and presents the found beta functions and fixed points.
We categorize the SMEFT operators in Sec.~\ref{sec:categories}, using our toy model as a paradigmatic example. We also find a mechanism for turning specific interactions into relevant ones based on these categories.
In Sec.~\ref{sec:light_fermions_take_5} we discuss the implications of our findings on the lightness of fermions in the Standard Model in view of previous investigations.
We conclude with Sec.~\ref{sec:conclusions}.

\section{Prerequisites}
\label{sec:prerequisites}

\subsection{Standard Model and Standard Model Effective Field Theory -- Field content and Lagrangian}

The SM describes quarks, leptons, Higgs and gauge fields and specifies charges under the symmetry group $\mathrm{SU}(3)_C\times\mathrm{SU}(2)_L\times\mathrm{U}(1)_Y$ (for charges see, e.g., \cite{Grzadkowski:2010es}).
The action governing the dynamics of the SM reads
\begin{equation}
    \begin{aligned}
	S_{\text{SM}} 
  = \int \!\mathrm{d}^4x \bigg\{& \!\!-\frac{1}{4} \!\left(G_{\mu\nu}^A G^{A\,\mu\nu} \!+ W_{\mu\nu}^I W^{I\,\mu\nu} \!+ B_{\mu\nu} B^{\mu\nu}\right) \\
      &+ \vphantom{\frac{1}{2}} \left(D_\mu \varphi\right)^\dagger \!\left(D^\mu \varphi\right) + m^2 \varphi^\dagger \varphi - \frac{\lambda}{2} \left(\varphi^\dagger \varphi\right)^2\\
				 &+ \vphantom{\frac{1}{2}} \ii \left(q^\dagger\! \slashed{D} q + u^\dagger\! \slashed{D} u + d^\dagger\! \slashed{D} d + l^\dagger\! \slashed{D} l + e^\dagger\! \slashed{D} e\right) \\
     &- \vphantom{\frac{1}{2}} \left(l^\dagger \Gamma^e e \varphi + q^\dagger \Gamma^u u \tilde{\varphi} + q^\dagger \Gamma^d d \varphi + \hc \right)\bigg\}\,.
    \end{aligned}
    \label{eq:SM_action}
\end{equation}
We use adjoint $\mathrm{SU}(3)_C$ indices $A \in \set{1,\dots,8}$ and adjoint $\mathrm{SU}(2)_L$ indices $I\in\set{1,2,3}$.
We denote $\tilde{\varphi}^a = \levicivita^{ab}\varphi^{*}_b$.
The covariant derivative is denoted $D$; $\slashed{D}$ abbreviates the contraction of the covariant derivative with the appropriate spin matrix, i.e., $\slashed{D}q = \barsigma^{\mu} D_\mu q$ for left-handed and $\slashed{D}u = \sigma^\mu D_\mu u$ for the right-handed case (see App.~\ref{app:weyl_fermions_in_qg} for details). 
Generation indices are suppressed, so the Yukawa couplings $\Gamma$ should be interpreted as matrices in generation space.

\afterpage{
\begin{table}[!h] 
\centering
\scriptsize
\renewcommand{\arraystretch}{1.5}
\begin{tabular}{lc|lc|lc} 

\multicolumn{2}{c}{$X^3$} &
\multicolumn{2}{|c|}{$\varphi^6$~ and~ $\varphi^4 D^2$} &
\multicolumn{2}{c}{$\psi^2\varphi^3$} \\
\hline
$O_G$ & $f^{ABC} G_\mu^{A\nu} G_\nu^{B\rho} G_\rho^{C\mu} $ & $O_\varphi$ & $(\varphi^\dag\varphi)^3$ & $O_{e\varphi}$ & $(\varphi^\dagger \varphi)(l_p^\dagger e_r \varphi)+\hc$\\
$O_{\tilde G}$ & $f^{ABC} \tilde G_\mu^{A\nu} G_\nu^{B\rho} G_\rho^{C\mu} $ & $O_{\varphi\Box}$ & $(\varphi^\dagger \varphi)\Box(\varphi^\dagger \varphi)$ & $O_{u\varphi}$ & $(\varphi^\dagger \varphi)(q_p^\dagger u_r \tilde{\varphi})+\hc$  \\
$O_W$ & $\levicivita^{IJK} W_\mu^{I\nu} W_\nu^{J\rho} W_\rho^{K\mu}$ & $O_{\varphi D}$   & $\left(\varphi^\dagger D^\mu\varphi\right)^* \!\left(\varphi^\dagger D_\mu\varphi\right)$ &
$O_{d\varphi}$ & $(\varphi^\dagger \varphi)(q_p^\dagger d_r \varphi)+\hc$\\
$O_{\tilde W}$ & $\levicivita^{IJK} \tilde W_\mu^{I\nu} W_\nu^{J\rho} W_\rho^{K\mu}$ &&&&
\vspace{10pt}
\end{tabular}

\begin{tabular}{lc|lc|lc} 
\multicolumn{2}{c}{$X^2\varphi^2$} &
\multicolumn{2}{|c|}{$\psi^2 X\varphi$} &
\multicolumn{2}{c}{$\psi^2\varphi^2 D$} \\ 
\hline
$O_{\varphi G}$ & $\varphi^\dagger \varphi\, G^A_{\mu\nu} G^{A\mu\nu}$ & $O_{eW}$ & $(l_p^\dagger \barsigma^{\mu\nu} e_r) \tau^I \varphi W_{\mu\nu}^I+\hc$ & $O_{\varphi l}^{(1)}$ & $\ii(\varphi^\dagger\Dleftright\varphi)(l_p^\dagger \barsigma^\mu l_r)$ \\
$O_{\varphi\tilde G}$ & $\varphi^\dagger \varphi\, \tilde G^A_{\mu\nu} G^{A\mu\nu}$ & $O_{eB}$ & $(l_p^\dagger \barsigma^{\mu\nu} e_r) \varphi B_{\mu\nu}+\hc$ & $O_{\varphi l}^{(3)}$ & $\ii(\varphi^\dagger\DIleftright\varphi)(l_p^\dagger \tau^I \barsigma^\mu l_r)$ \\
$O_{\varphi W}$ & $\varphi^\dagger \varphi\, W^I_{\mu\nu} W^{I\mu\nu}$ & $O_{uG}$ & $(q_p^\dagger \barsigma^{\mu\nu} T^A u_r) \tilde{\varphi}\, G_{\mu\nu}^A+\hc$ & $O_{\varphi e}$ & $\ii(\varphi^\dagger\Dleftright\varphi)(e_p^\dagger \sigma^\mu e_r)$ \\
$O_{\varphi\tilde W}$ & $\varphi^\dagger \varphi\, \tilde W^I_{\mu\nu} W^{I\mu\nu}$ & $O_{uW}$ & $(q_p^\dagger \barsigma^{\mu\nu} u_r) \tau^I \tilde{\varphi}\, W_{\mu\nu}^I+\hc$ & $O_{\varphi q}^{(1)}$ & $\ii(\varphi^\dagger\Dleftright\varphi)(q_p^\dagger \barsigma^\mu q_r)$\\
$O_{\varphi B}$ & $ \varphi^\dagger \varphi\, B_{\mu\nu} B^{\mu\nu}$ & $O_{uB}$ & $(q_p^\dagger \barsigma^{\mu\nu} u_r) \tilde{\varphi}\, B_{\mu\nu}+\hc$ & $O_{\varphi q}^{(3)}$ & $\ii(\varphi^\dagger\DIleftright\varphi)(q_p^\dagger \tau^I \barsigma^\mu q_r)$ \\
$O_{\varphi\tilde B}$ & $\varphi^\dagger \varphi\, \tilde B_{\mu\nu} B^{\mu\nu}$ & $O_{dG}$ & $(q_p^\dagger \barsigma^{\mu\nu} T^A d_r) \varphi\, G_{\mu\nu}^A+\hc$ & $O_{\varphi u}$ & $\ii(\varphi^\dagger\Dleftright\varphi)(u_p^\dagger \sigma^\mu u_r)$ \\
$O_{\varphi WB}$ & $ \varphi^\dagger \tau^I \varphi\, W^I_{\mu\nu} B^{\mu\nu}$ & $O_{dW}$ & $(q_p^\dagger \barsigma^{\mu\nu} d_r) \tau^I \varphi\, W_{\mu\nu}^I+\hc$ & $O_{\varphi d}$ & $\ii(\varphi^\dagger\Dleftright\varphi)(d_p^\dagger \sigma^\mu d_r)$ \\
$O_{\varphi\tilde WB}$ & $\varphi^\dagger \tau^I \varphi\, \tilde W^I_{\mu\nu} B^{\mu\nu}$ & $O_{dB}$ & $(q_p^\dagger \barsigma^{\mu\nu} d_r) \varphi\, B_{\mu\nu}+\hc$ & $O_{\varphi u d} $ & $\ii(\tilde{\varphi}^\dagger D_\mu \varphi)(u_p^\dagger \sigma^\mu d_r)+\hc$
\vspace{10pt}
\end{tabular}

\begin{tabular}{cc|cc|cc}
\multicolumn{2}{c}{$(L^\dagger\! L)(L^\dagger\! L)$} & 
\multicolumn{2}{|c|}{$(R^\dagger\! R)(R^\dagger\! R)$} &
\multicolumn{2}{c}{$(L^\dagger\! L)(R^\dagger\! R)$} \\
\hline
$O_{ll}$& $\underline{(l_p^\dagger \barsigma_\mu l_r)(l_s^\dagger \barsigma^\mu l_t)}$ & $O_{ee}$ & $\underline{(e_p^\dagger \sigma_\mu e_r)(e_s^\dagger \sigma^\mu e_t)}$ & $O_{le}$ & $\underline{(l_p^\dagger \barsigma_\mu l_r)(e_s^\dagger \sigma^\mu e_t)}$ \\
$O_{qq}^{(1)}$ & $\underline{(q_p^\dagger \barsigma_\mu q_r)(q_s^\dagger \barsigma^\mu q_t)}$ & $O_{uu}$ & $\underline{(u_p^\dagger \sigma_\mu u_r)(u_s^\dagger \sigma^\mu u_t)}$ & $O_{lu}$ & $\underline{(l_p^\dagger \barsigma_\mu l_r)(u_s^\dagger \sigma^\mu u_t)}$ \\
$O_{qq}^{(3)}$ & $(q_p^\dagger \barsigma_\mu \tau^I q_r)(q_s^\dagger \barsigma^\mu \tau^I q_t)$ & $O_{dd}$ & $\underline{(d_p^\dagger \sigma_\mu d_r)(d_s^\dagger \sigma^\mu d_t)}$ &
$O_{ld}$ & $\underline{(l_p^\dagger \barsigma_\mu l_r)(d_s^\dagger \sigma^\mu d_t)}$ \\
$O_{lq}^{(1)}$ & $\underline{(l_p^\dagger \barsigma_\mu l_r)(q_s^\dagger \barsigma^\mu q_t)}$ & $O_{eu}$ & $\underline{(e_p^\dagger \sigma_\mu e_r)(u_s^\dagger \sigma^\mu u_t)}$ & $O_{qe}$ & $\underline{(q_p^\dagger \barsigma_\mu q_r)(e_s^\dagger \sigma^\mu e_t)}$ \\
$O_{lq}^{(3)}$ & $(l_p^\dagger \barsigma_\mu \tau^I l_r)(q_s^\dagger \barsigma^\mu \tau^I q_t)$ & $O_{ed}$ & $\underline{(e_p^\dagger \sigma_\mu e_r)(d_s^\dagger \sigma^\mu d_t)}$ & $O_{qu}^{(1)}$ & $\underline{(q_p^\dagger \barsigma_\mu q_r)(u_s^\dagger \sigma^\mu u_t)}$ \\ 
&& $O_{ud}^{(1)}$ & $\underline{(u_p^\dagger \sigma_\mu u_r)(d_s^\dagger \sigma^\mu d_t)}$ & $O_{qu}^{(8)}$ & $(q_p^\dagger \barsigma_\mu T^A q_r)(u_s^\dagger \sigma^\mu T^A u_t)$ \\ 
&& $O_{ud}^{(8)}$ & $(u_p^\dagger \sigma_\mu T^A u_r)(d_s^\dagger \sigma^\mu T^A d_t)$ & $O_{qd}^{(1)}$ & $\underline{(q_p^\dagger \barsigma_\mu q_r)(d_s^\dagger \sigma^\mu d_t)}$ \\
&&&& $O_{qd}^{(8)}$ & $(q_p^\dagger \barsigma_\mu T^A q_r)(d_s^\dagger \sigma^\mu T^A d_t)$
\vspace{10pt}
\end{tabular}

\begin{tabular}{cc|cc}
\multicolumn{2}{c|}{$(L^\dagger\! R)(R^\dagger\! L)$ and $(L^\dagger\! R)(L^\dagger\! R)$} & \multicolumn{2}{c}{$B$-violating} \\\hline
$O_{ledq}$ & $(l_p^{\dagger a} e_r)(d_s^\dagger q_t^a) +\hc$ & $O_{duq}$ & {$\levicivita^{\alpha\beta\gamma} \levicivita_{ab} \Big[ (d^\alpha_p)^T C u^\beta_r \Big]\Big[(q^{\gamma a}_s)^T C l^b_t\Big] +\hc$}\\
$O_{quqd}^{(1)}$ & $(q_p^{\dagger a} u_r) \levicivita_{ab} (q_s^{\dagger b} d_t) +\hc$ & $O_{qqu}$ & {$\levicivita^{\alpha\beta\gamma} \levicivita_{ab} \Big[ (q^{\alpha a}_p)^T C q^{\beta b}_r \Big]\Big[(u^\gamma_s)^T C e_t\Big] +\hc$} \\
$O_{quqd}^{(8)}$ & $(q_p^{\dagger a} T^A u_r) \levicivita_{ab} (q_s^{\dagger b} T^A d_t)+\hc$ & $O_{qqq}$ & {$\levicivita^{\alpha\beta\gamma} \levicivita_{ad} \levicivita_{bc} \Big[ (q^{\alpha a}_p)^T C q^{\beta b}_r \Big]\Big[(q^{\gamma c}_s)^T C l^d_t\Big] +\hc$} \\
$O_{lequ}^{(1)}$ & $(l_p^{\dagger a} e_r) \levicivita_{ab} (q_s^{\dagger b} u_t)+\hc$ & $O_{duu}$ & {$\levicivita^{\alpha\beta\gamma} \Big[ (d^\alpha_p)^T C u^\beta_r \Big]\Big[(u^\gamma_s)^T C e_t\Big]+\hc$} \\
$O_{lequ}^{(3)}$ & $(l_p^{\dagger a} \barsigma_{\mu\nu} e_r) \levicivita_{ab} (q_s^{\dagger b} \barsigma^{\mu\nu} u_t)+\hc$ & & {} \\
\end{tabular}
\caption{
    SMEFT mass-dimension-six operators.
    The upper two panels describe the gauge and Higgs sector whereas the lower panels list four-fermion interactions.
    The underlined four-fermion operators respect the symmetries of the free theory.
    They must be non-zero, but Planck-scale suppressed. 
    On the other hand, non-underlined operators may become relevant and take a non-zero and not Planck-scale suppressed value in the IR, see Sec.~\ref{sec:categories} for details. 
}
\label{tab:smeft_dim_6}
\end{table}
}

The SMEFT uses the SM field content and considers all (quasi-)local operators that respect Lorentz invariance and the SM gauge group $\mathrm{SU}(3)_C \times \mathrm{SU}(2)_L \times \mathrm{U}(1)_Y$ at any mass dimension \cite{Grzadkowski:2010es}.
At mass dimension $\le 4$, this implies the SM action \eqref{eq:SM_action}.\footnote{In addition, one could in principle write down a $\theta_{\text{QCD}}$-term. However, it is a total 4-divergence -- its variation vanishes, so it does not renormalize the operators we study here.}
The higher-order operators are considered to be effective interactions induced by new physics above a UV scale $\Lambda_{\text{UV}}$.
The full action is 
\begin{equation}
	S_{\text{SMEFT}} = S_{\text{SM}} + \sum_{D>4}^{\infty} \sum_{i}\frac{c_i}{\Lambda_{\text{UV}}^{D-4}}\int \mathrm{d}^4 x \, O_{i,D} \,.
\end{equation}
In the absence of a concrete UV completion, one can make the naturalness assumption that the dimensionless coefficients $c_i$ are of order unity.
We may focus on the mass-dimension-six operators and neglect higher order contributions for the dominant phenomenological effects, e.g., at the LHC.

At mass-dimension 6 there are $84$ structurally independent hermitian operators in the SMEFT\footnote{This count includes 8 baryon-number-violating terms.} \cite{Henning:2015alf}.
They are listed in Tab.~\ref{tab:smeft_dim_6} in the Warsaw Basis \cite{Grzadkowski:2010es}.
$38$ of the mass-dimension-six operators are four-fermion operators \cite{Grzadkowski:2010es}.
For the systematic construction of SMEFT operators under consideration of ambiguities arising from Fierz identities and integration by parts / EOM ambiguities see \cite{Grzadkowski:2010es, Henning:2015alf}. 
In this table, $f^{ABC}$ are the structure constants and $T^A$ are the generators of $\mathrm{SU}(3)_C$.
The dual field strength tensors are defined as $\tilde{X}_{\mu\nu} = \frac{1}{2} \levicivita_{\mu\nu\rho\sigma} X^{\rho\sigma}$.
For $\mathrm{SU}(2)_L$, the Levi-Civita symbol $\levicivita^{IJK}$ assumes the role of the structure constants and the generators are $\tau^I$.
$\hc$ abbreviates the hermitian conjugate.
$p,r,s,t$ are generation indices.
The sigma matrices with two Lorentz indices are defined as $\sigma^{\mu\nu} = \frac{\ii}{4}\left(\sigma^{\mu} \barsigma^{\nu} - \sigma^{\nu} \barsigma^{\mu}\right)$ and $\barsigma^{\mu\nu} = \frac{\ii}{4}\left(\barsigma^{\mu} \sigma^{\nu} - \barsigma^{\nu} \sigma^{\mu}\right)$.
The operator $\;\Dleftright\;$ acts to the left and the right within parentheses $(\varphi^\dagger \Dleftright \varphi) = (\varphi^\dagger D_\mu \varphi - (D_\mu \varphi^\dagger) \varphi)$.
The version with an adjoint $\mathrm{SU}(2)_L$ index includes a generator $(\varphi^\dagger \DIleftright \varphi) = (\varphi^\dagger \tau^I D_\mu \varphi - (D_\mu \varphi^\dagger) \tau^I\varphi)$.
$a,b,c,d$ are isospin indices and $\alpha,\beta,\gamma$ are color indices.
In the Baryon number violating sector, $C$ is the charge conjugation operator.
For details on the notation see \cite{Grzadkowski:2010es}.

In Tab.~\ref{tab:smeft_dim_6}, we distinguish two categories of operators: Operators which are underlined  necessarily have non-zero SMEFT coefficients in asymptotically safe quantum gravity.
We explain the underlying mechanism in Sec.~\ref{sec:categories}.
In contrast, operators which are not underlined, may have vanishing SMEFT coefficients.
We stress that even non-zero SMEFT coefficients may still be Planck-scale suppressed and thus unmeasurably small for practical purposes.
The classification into underlined and non-underlined operators is to be understood as referring to \emph{structural} aspects of the SMEFT with gravity, not phenomenological aspects.
Finally, in Sec.~\ref{sec:IR-regime}, we will find that the non-underlined operators can potentially become free parameters, such that Planck-scale suppression is not guaranteed.
Instead, they may deviate from zero in some scenarios, and it is unclear whether they respect the EFT ordering.
In contrast, this mechanism does not apply to the underlined SMEFT coefficients which are necessarily non-zero in the UV.
These generically respect the EFT ordering and corresponding Planck-scale suppression, see Sec.~\ref{sec:UV-regime}.

\FloatBarrier

\subsection{Background on asymptotically safe quantum gravity}

Asymptotically safe gravity, first proposed by Weinberg \cite{Weinberg:1980gg}, is based on a fixed point of the Renormalization Group (RG) flow. Such a fixed point relies on quantum fluctuations of the metric field. It addresses the perturbative non-renormalizability of quantum gravity and underlies a quantum field theory of the metric.

\subsubsection{Fixed point and origin of predictivity}

Perturbative renormalization of quantum gravity produces infinitely many counterterms. At loop order $L$, terms with $2(L+1)$ derivatives arise. These are organized into higher-order curvature terms that have to be added to the action in addition to the Einstein-Hilbert term.

Asymptotic safety adds one additional requirement, namely that the coefficients of these terms are constrained by an infinite tower of relations. These relations arise when the RG flow of the coefficients starts from a fixed point in the UV.

In short, rather than starting from arbitrary initial conditions $g_i(k_{\rm UV})$ for all (dimensionless) couplings $g_i(k)$ at some scale $k_{\rm UV}$, one starts from
\begin{equation}
g_i(k_{\rm UV}) = g_{i\, \ast} + \delta^{(\rm rel.)} g_i\,.
\end{equation}
In here, $g_{i\, \ast}$ is the fixed-point value, i.e., the beta function $\beta_{g_i} = k\partial_k\, g_i(k)$ vanishes at this point
\begin{equation}
    \beta_{g_i}\Big|_{\vec{g}=\vec{g}_{\ast}}=0 \quad\forall \beta_{g_i}\,,
\end{equation}
where we have summarized all couplings into a vector $\vec{g}$. $\delta^{\rm (rel.)}g_i$ is a relevant perturbation of the fixed point. In other words, it is a direction in coupling space, along which the RG flow towards the IR is directed away from the fixed point rather than towards it. 

The distinction between relevant directions and their counterparts, irrelevant directions, can be made on the basis of the eigenvalues of the stability matrix
\begin{equation}
\theta_I = - {\rm eig} \left(\frac{\partial \beta_{g_i}}{\partial g_j} \right)\Big|_{\vec{g}= \vec{g}_{\ast}}.
\end{equation}
If $\theta_I>0$, then the corresponding eigenvector in the space of couplings constitutes a relevant perturbation; if $\theta^I<0$, it constitutes an irrelevant perturbation. The RG flow is driven back into the fixed point along the irrelevant directions. Therefore, these directions are automatically constrained at all scales. Conversely, the RG flow towards the IR can leave the fixed point along relevant directions, such that each relevant direction gives rise to a free parameter of the effective action in the IR.

Ultimately, the effective action that describes an asymptotically safe theory has infinitely many operators, whose couplings are parameterized by a finite number of free parameters. These free parameters encode the distances from the fixed point along the distinct relevant directions. 

Applied concretely to the SMEFT, we expect that its UV completion with asymptotically safe quantum gravity provides constraints on the SMEFT coefficients associated to the irrelevant directions of the fixed point.

\subsubsection{Status of asymptotic safety in gravity-matter systems}

Following Weinberg's original proposal \cite{Hawking:1979ig}, some work was done on asymptotic safety in the epsilon-expansion about two dimensions, see \cite{Martini:2022sll} for a recent review, but did not lead to conclusive results for the case of interest, namely four spacetime dimensions. The situation changed with Reuter's seminal paper \cite{Reuter:1996cp} which pioneered functional RG techniques, see Sec.~\ref{sec:FRG}, for gravity. Based on a very large body of literature, the existence of an asymptotically safe fixed point in Euclidean, pure gravity is now regarded as established, see  \cite{Niedermaier:2006wt,Reuter:2012id,Eichhorn:2018yfc,Pereira:2019dbn,Pawlowski:2020qer,Eichhorn:2022jqj,Eichhorn:2022gku,Eichhorn:2022bgu,Wetterich:2022ncl,Knorr:2022dsx,Morris:2022btf,Martini:2022sll,Pawlowski:2023gym,Saueressig:2023irs,Platania:2023srt,Bonanno:2024xne,Eichhorn:2023xee} for reviews, with \cite{Eichhorn:2026uqj} an up-to-date review. Supporting evidence also comes from lattice simulations \cite{Ambjorn:2024qoe}. 

A central property of this fixed point emerged from the studies in \cite{Falls:2013bv, Falls:2014tra, Falls:2017lst, Falls:2018ylp,Kluth:2020bdv,Kluth:2022vnq}, as well as \cite{Eichhorn:2018akn, Eichhorn:2018ydy,Eichhorn:2018nda, Eichhorn:2020sbo} and most recently complemented by \cite{Assant:2026dca}. It consists in the fact that the deviation of the critical exponents $\theta_I$ from the canonical scaling dimensions of couplings is not large, except for the scaling dimension of the Newton coupling. This entails that the EFT ordering into relevant and irrelevant couplings, which is strictly correct at the free fixed point, remains a very good guiding principle at the interacting fixed point. This crucial property means that truncations, inherently necessary in the use of functional RG techniques, follow an ordering principle, so that operators that are neglected can be relied on to actually be irrelevant.

Steps towards a more realistic setting than pure, Euclidean gravity have been undertaken over the last decade. Lorentzian signature has been studied in \cite{Fehre:2021eob,DAngelo:2022vsh,DAngelo:2023tis,Pawlowski:2025etp,Kher:2025rve,Assant:2026dca}, supported by results in Euclidean signature that admit an analytical continuation \cite{Manrique:2011jc,Rechenberger:2012dt,Biemans:2016rvp,Houthoff:2017oam,Biemans:2017zca,Saueressig:2023tfy,Korver:2024sam,Saueressig:2025ypi,Bonanno:2021squ,Pastor-Gutierrez:2024sbt,Chiesa:2026tlz,Knorr:2026jcg}. At this moment, the subset of gravitational operators that has been included in these studies is largely limited to the Einstein-Hilbert truncation. However, the similarity of the resulting RG flows to those in Euclidean signature is promising and may hint that the change in signature only produces subleading effects.

 As a distinct step towards a more realistic setting, matter degrees of freedom have been included, see \cite{Eichhorn:2022jqj,Eichhorn:2022gku, Eichhorn:2026uqj} for reviews, resulting in the conclusion that an asymptotically safe fixed point persists under the inclusion of SM matter \cite{Dona:2013qba}, see also  \cite{Meibohm:2015twa, Biemans:2017zca, Christiansen:2017cxa,Alkofer:2018fxj, Wetterich:2019zdo, Pastor-Gutierrez:2022nki}. In addition, predictive power for SM couplings was established in \cite{Shaposhnikov:2009pv,Harst:2011zx, Eichhorn:2017lry, Eichhorn:2017ylw, Eichhorn:2018whv}. First papers even study Lorentzian gravity-matter systems \cite{Pastor-Gutierrez:2024sbt,Kher:2025rve,Chiesa:2026tlz}, with largely promising first results, although it is too early to make robust statements about the viability of the asymptotically safe SM in Lorentzian signature.
 
 Beyond the dimension-four-interactions of the SM, various higher-order interactions, structurally similar to SMEFT operators, although typically not for the SM gauge group, have been studied \cite{Narain:2009fy,Eichhorn:2011pc,Eichhorn:2012va,Percacci:2015wwa,Meibohm:2016mkp,Eichhorn:2016vvy,Eichhorn:2016esv,Christiansen:2017gtg,Eichhorn:2017eht,Hamada:2017rvn,Eichhorn:2017sok,Eichhorn:2019yzm,deBrito:2020dta,Daas:2020dyo,Daas:2021abx,Laporte:2021kyp,deBrito:2021pyi,Eichhorn:2021qet,Eichhorn:2022ngh,deBrito:2023myf, deBrito:2023kow, Eichhorn:2023jyr, Brenner:2024bps, deBrito:2025nog}.
 These interactions falls into two categories, namely those with vanishing fixed-point value \cite{Narain:2009fy,Percacci:2015wwa,Eichhorn:2016vvy,Hamada:2017rvn,Daas:2020dyo,Daas:2021abx,Eichhorn:2023jyr,deBrito:2025nog} as well as non-vanishing fixed-point value \cite{Eichhorn:2011pc,Eichhorn:2012va,Meibohm:2016mkp,Eichhorn:2016esv,Christiansen:2017gtg,Eichhorn:2017eht,Eichhorn:2017sok,Eichhorn:2019yzm,deBrito:2020dta,Laporte:2021kyp,deBrito:2021pyi,Eichhorn:2021qet,Eichhorn:2022ngh,deBrito:2023myf, deBrito:2023kow, Brenner:2024bps,deBrito:2025nog}.
 The reason for this difference is the role that global symmetries play in asymptotic safety.

\subsubsection{Global symmetries and induced interactions in asymptotically safe quantum gravity}
\label{sec:globalsymmetries}

Matter is never entirely non-interacting in the presence of quantum gravity.
This is because quantum gravity fluctuations couple to the kinetic terms for matter fields and can thus mediate interactions.
Within asymptotic safety, this implies that graviton loops give rise to matter self-interactions that cannot be set to zero at the asymptotically safe fixed point \cite{Eichhorn:2011pc, Eichhorn:2012va}. The size of these gravitationally induced interactions is parametrically controlled by the gravitational coupling $g$. Thus, a fixed point at which $g_{\ast}\neq 0$ cannot have vanishing interactions for the SM fields.
Rather, these are fewer interactions, subject to a larger global symmetry group, namely the maximal symmetry of the kinetic terms when all gauge couplings are set to zero.\footnote{Non-Abelian gauge couplings retain asymptotic freedom under the impact of quantum gravity \cite{Daum:2009dn,Folkerts:2011jz} and therefore vanish at the fixed point.
    These are, however, not all matter interactions that are compatible with the gauge symmetries of the matter model.
For the Abelian gauge coupling, there are two fixed points, a free and an interacting one \cite{Harst:2011zx, Eichhorn:2017lry}.
If the interacting fixed point takes precisely the correct value to reproduce the value of the Abelian hypercharge coupling in the IR, it is the fixed point which must be used for a realistic asymptotically safe theory.
Otherwise, the interacting fixed point results in an upper bound and the Abelian hypercharge also becomes asymptotically free and therefore vanishes at the fixed point.
In our discussion, we assume this latter scenario.
If instead, the interacting fixed point is realized, our discussion generalizes, see \cite{Eichhorn:2017eht}, and the global symmetry group that dictates the properties of the gravity-matter fixed point is different and accounts for non-vanishing hypercharges of matter fields.}
Inspecting the class of such interactions, we note that their dimensionality is at least six (for fermions \cite{Eichhorn:2011pc, Meibohm:2015twa, Eichhorn:2017eht, deBrito:2020dta, deBrito:2023kow}) and eight (for scalars \cite{Eichhorn:2012va, deBrito:2021pyi,Laporte:2021kyp, deBrito:2023myf} and gauge fields \cite{Christiansen:2017gtg, Eichhorn:2021qet, Eichhorn:2024wba, Knorr:2024yiu} as well as any interactions that mix different fields \cite{Eichhorn:2016esv}).\footnote{ Additionally, there are induced non-minimal  interactions \cite{Eichhorn:2017sok, Laporte:2021kyp, Knorr:2024yiu,deBrito:2025nog}, but these are expected to be Planck-scale suppressed, in addition to vanishing on a flat background.} Combined with the near-perturbative scaling spectrum of the asymptotically safe fixed point, this implies that we generically expect these interactions to be irrelevant at the combined gravity-matter fixed point. 

Such a fixed point, at which only the gravitationally induced interactions exist for matter, has been called \emph{maximally symmetric asymptotically safe fixed point} (MSAS) in \cite{Eichhorn:2017eht}.

It is not a given that this fixed point exists. From the reasoning above, we can only conclude that the \emph{Gaussian fixed point for matter does not exist}. Thus, we can hope for a \emph{shifted} Gaussian fixed point (SGFP), the MSAS fixed point, for which all couplings compatible with the symmetries of the kinetic terms are shifted to nonzero values, parametrically controlled by $g$.
For a recent compilation of interactions induced by gravity in the asymptotic regime see \cite{Eichhorn:2022gku}.

We note the special role that global symmetries play in this discussion. Their preservation at the SGFP shapes the properties of this fixed point. 
In addition, global symmetries shape the resulting effective action that one obtains by integrating over the RG scale to the IR, starting from the asymptotically safe fixed point. Because the RG flow does not violate a symmetry that is respected by the initial condition given an appropriate regulator, no symmetry-violating interactions arise.\footnote{Symmetry-violating interactions can be non-zero in the IR, if they are connected to a relevant perturbation of the fixed point. Then, it is a possibility (but not a necessity) for such an interaction to be nonzero in the IR.}
This may seem to violate the no-global-symmetries-conjecture formulated in the context of string theory for quantum gravity \cite{Banks:1988yz, Banks:2010zn}, but expected by many to hold for any consistent quantum theory of gravity, see, however \cite{Eichhorn:2024rkc,Basile:2025zjc}.
This conjecture is based on the intuition that black holes  are able to destroy infalling information and thus attached Noether charges from global symmetries. 
If black-hole entropy in the quantum gravity regime deviates from Hawking's formula \cite{Basile:2025zjc}, or if black-hole configurations are dynamically suppressed in the path integral \cite{Borissova:2020knn, Borissova:2023kzq, Borissova:2020knn}, the conjecture no longer applies. 
There is, however, also the possibility that calculations in asymptotic safety do not properly account for black-hole configurations, because they are largely conducted in Euclidean signature, where no black-hole interiors are included among the field configurations.

Yet, we stress that the underlying symmetry group $\mathrm{SU}(2)_L\times\mathrm{U}(1)_Y$ we use as an input is fundamentally gauged, such that the no-global-symmetries-conjecture does not apply to the actual system we are interested in.
Moreover, the maximal symmetry that the SGFP fulfills is not given as an assumption but rather emerges dynamically from asymptotic scale symmetry. Therefore for the remainder of this paper, we work under the assumption that the role of global symmetries in determining the fixed-point properties is reliably captured by our calculations.

\subsection{Methodology: Functional Renormalization Group}
\label{sec:FRG}

To explore the behavior of SMEFT operators in asymptotic safety, we employ the functional renormalization group (FRG) \cite{Wetterich:1992yh, Morris:1993qb, Ellwanger:1993mw}.
At its core is the Wetterich equation
\begin{equation}
	\label{eq:wetterich}
	\begin{aligned}
		k \partial_k \Gamma_k
		 = \frac{1}{2} \STr \left(\big(\Gamma^{(2)}_k + R_k\big)^{-1} \cdot k \partial_k R_k\right).
	\end{aligned}
\end{equation}
It describes the gradual integration of quantum fluctuations in the path integral. Herein, $\Gamma_k$ is a scale-dependent effective action. In the limit $k \rightarrow 0$, it agrees with the standard effective action, from which one can extract the Wilson coefficients. At $k>0$, $\Gamma_k$ differs from $\Gamma$ by the effect of long-wavelength quantum fluctuations. In the underlying path integral, fluctuations are partitioned into momentum modes, where the renormalization group scale $k$ characterizes the momentum shell.
The infrared regulator $R_k (p)$ is chosen to have the tensor structure of the kinetic operator and it includes a shape function $r_k(p^2)$ that controls which momenta are included and which are absent. It is non-zero for $p^2<k^2$ and resembles a mass term for these modes. This results in a suppression in the path integral. Conversely, the regulator vanishes for $p^2>k^2$, such that the effect of those modes is accounted for in $\Gamma_k$.

The Wetterich equation describes the exact renormalization group evolution of the effective action $\Gamma_k$ via the supertrace of its Hessian $\Gamma_k^{(2)}$ and the regulator.
The supertrace includes an integration of internal momenta and supplements fermionic loops with a negative sign.
The loop-momentum-integral on the right-hand-side of the Wetterich equation receives the main contribution from modes with $p^2 \approx k^2$. Thus, the change of $\Gamma_k$ with $k$ is largely driven by quantum fluctuations at this scale.

In the context of the SMEFT, it is noteworthy that the Wetterich equation automatically accounts for the decoupling of massive modes. This is an advantage when constructing the low-energy effective theory that corresponds to a given UV completion, because no intricate matching prescriptions are needed. Instead, as the RG scale falls below the mass scale of a mode, $k^2<m^2$, the contribution of that mode to the RG flow is automatically suppressed. This makes it possible to extract Wilson coefficients without the need for decoupling prescriptions, see, e.g., \cite{Eichhorn:2024wba, Knorr:2024yiu} for examples in gravity.

For a review of the FRG and its applications see, e.g., \cite{Dupuis:2020fhh}. For reviews more specifically about the application of the FRG in gravity, see \cite{Saueressig:2023irs, Pawlowski:2023gym, Knorr:2022dsx}.

\section{Four-fermion interactions in the SMEFT with quantum gravity}

\subsection{Action and beta functions}
\label{sec:action}

To relate the SMEFT coefficients to asymptotically safe quantum gravity, we compute the RG evolution of a subset of SMEFT coefficients under the impact of quantum gravity fluctuations.\footnote{
RG studies of SMEFT coefficients without gravity can be found in \cite{Bartocci:2024fmm,DiNoi:2025tka,terHoeve:2025gey,DasBakshi:2026ief,Mantani:2026fao}.} From this subset, we then draw generalized lessons about all coefficients at dimension six.
We study a fermionic sector charged under $\mathrm{SU}(2)_L \times \mathrm{U}(1)_Y$, modeling one generation of quarks. 
The system is thus composed of one left-handed quark doublet $q$ and two right-handed quarks $u$ and $d$. 
Charges are considered global, and thus gauge fields are neglected. This is a good approximation at the fixed point at which all gauge couplings, including the $U(1)_Y$, can become asymptotically free under the impact of quantum gravity \cite{Daum:2009dn,Harst:2011zx,Folkerts:2011jz, Christiansen:2017cxa, Christiansen:2017gtg,Eichhorn:2017lry, Eichhorn:2019yzm}.
In addition, we suppress the $\mathrm{SU}(3)_C$ charge, i.e., we do not consider color indices of the quarks.

We consider the following sectors in our ansatz for the scale-dependent effective action
\begin{equation}
    \Gamma_k = \Gamma_k^{\text{4F}} + \Gamma_k^{\text{kin.}} + \Gamma_k^{\text{EH}} + \Gamma_k^{\text{g.f.}} + \Gamma_k^{\text{gh.}}  \,.
	\label{eq:total_action}
\end{equation}
Herein, $\Gamma_k^{\text{4F}}$ contains the four-fermion interactions and $\Gamma_k^{\text{kin.}}$ the kinetic terms for the corresponding fermions. $\Gamma_k^{\text{EH}}  + \Gamma_k^{\text{g.f.}} + \Gamma_k^{\text{gh.}}$ contains the gravitational dynamics in a gauge-fixed formulation, i.e., $\Gamma_k^{\text{g.f.}}$ is the gauge-fixing and $\Gamma_k^{\text{gh.}}$ the corresponding Faddeev-Popov term.
For further details on the truncation see App.~\ref{app:general_setup}.

We start by considering the four-fermion interactions.
The chosen field content allows the four-fermion operators $O_{qq}^{(1)}$, $O_{uu\vphantom{d}}^{\vphantom{(1)}}$, $O_{dd}^{\vphantom{(1)}}$, $O_{ud}^{(1)}$, $O_{qu\vphantom{d}}^{(1)}$, $O_{qd}^{(1)}$ and $O_{quqd}^{(1)}$ of the $(L^\dagger\! L)(L^\dagger\! L)$, $(R^\dagger\! R)(R^\dagger\! R)$, $(L^\dagger\! L)(R^\dagger\! R)$ and $(L^\dagger\! R)(L^\dagger\! R)$ types.
As presented in Tab.~\ref{tab:smeft_dim_6}, these operators are given in the complex basis, where both the operators and couplings may be complex.
For our purposes, it is convenient to switch to the hermitian basis where both quantities are real.
For our truncation, this switch amounts to
\begin{equation}
    \begin{aligned}
    \big[ C_{quqd} O_{quqd} + \hc \big]
    &= C_{quqd} \,\levicivita^{ab}(q^{\dagger}_a u) (q^{\dagger}_b d) + (C_{quqd})^* \levicivita^{ab}(u^{\dagger} q_a) (d^{\dagger} q_b) \\
    &=  \CRe \big[ O_{quqd} + O_{quqd}^\dagger \big] + \ii\CIm \big[ O_{quqd} - O_{quqd}^\dagger \big],
    \end{aligned}
\end{equation}
where $\CRe = (C_{quqd} + C_{quqd}^*)/2 \in \mathds{R}$ and $\CIm = -\ii (C_{quqd} - C_{quqd}^*)/2 \in \mathds{R}$.

The fermion sector of the action, containing all eight\footnote{The operator $O_{qq}^{(3)} = (q_a^\dagger \barsigma_\mu \tau_I q^a)(q_b^\dagger \barsigma^\mu \tau^I q^b)$ with $\tau^I$ the generators of $\mathrm{SU}(2)_L$, is a dependent operator given our symmetries. It can be transformed into $(q_a^\dagger \barsigma_\mu q^a)(q_b^\dagger \barsigma^\mu q^b)$ using Fierz identities. When including $\mathrm{SU}(3)_C$ also, it becomes independent and must be considered with its own Wilson coefficient in the truncation.} 
four-fermion operators that make up a Fierz-complete set \cite{Grzadkowski:2010es, Henning:2015alf} then reads
\begin{align}
	\label{eq:4_fermion_action}
		\Gamma_{k}^{4F}
    = \int \dd[4]{x\sqrt{g}} \bigg\{
        &\phantom{+} C_{qq} (q^{\dagger}_a \barsigma_\mu q^a)(q^{\dagger}_b \barsigma^\mu q^b)
		+ C_{uu} (u^\dagger \sigma_\mu u) (u^\dagger \sigma^\mu u)
		+ C_{dd} (d^\dagger \sigma_\mu d)(d^\dagger \sigma^\mu d) \nn\\
    & + C_{qu} (q^{\dagger}_a \barsigma_\mu q^a)(u^\dagger \sigma^\mu u)
		+ C_{qd} (q^{\dagger}_a \barsigma_\mu q^a) (d^\dagger \sigma^\mu d)
    + C_{ud} (u^\dagger \sigma_\mu u)(d^\dagger \sigma^\mu d) \nn\\
& + \CRe \big[(\levicivita^{ab}(q^{\dagger}_a u) (q^{\dagger}_b d) + \hc\big] + \CIm \big[\levicivita^{ab}(q^{\dagger}_a u) (q^{\dagger}_b d) - \hc\big] \bigg\}\,.
\end{align}
Uppercase letters indicate dimensionful couplings and lowercase letters represent their dimensionless counterparts, such that 
\begin{equation}
C_i = c_i / k^2.
\end{equation}
In the hermitian formulation of this action we employed the choice of basis, which constitutes an $\mathrm{O}(2)$ rotation between the couplings $(C_{quqd}, C_{quqd*})$ and $(O_{quqd},O_{quqd}^\dagger)$.
    Because the other couplings are unaffected by this rotation, we expect their beta functions to depend only on the combination $(\CRe{})^2+(\CIm{})^2$.\footnote{This can be understood as an independence from phase of the complex coupling in the original complex representation.}

Additionally, there are three kinetic terms, one each for $q$, $u$ and $d$, given by
\begin{equation}
	\label{eq:fermion_kinetic_action}
	\begin{aligned}
		\Gamma_k^{\mathrm{kin.}}
		 & = \int\dd[4]{x \sqrt{g}} \Big\{
			\ii q^{\dagger}_a \barsigma^\mu \nabla_\mu q^a
			+ \ii u^{\dagger} \sigma^\mu \nabla_\mu u
		+ \ii d^{\dagger} \sigma^\mu \nabla_\mu d\Big\}\,.
	\end{aligned}
\end{equation}
We neglect the running of the field renormalization.
The spacetime-dependent sigma matrices $(\barsigma_{\mu})^{\dot{\alpha}\beta}(x)$, $(\sigma^{\mu})_{\alpha \dot{\beta}}(x)$ fulfill the identities
\begin{equation}
	\label{eq:pauli_clifford}
	\left(\sigma^\mu \barsigma^\nu + \sigma^\nu \barsigma^\mu\right)_\alpha^{\phantom{\alpha}\beta} = 2g^{\mu\nu}\delta_\alpha^{\phantom{\alpha}\beta}\,,\quad 
    \left(\barsigma^\mu \sigma^\nu + \barsigma^\nu \sigma^\mu\right)^\dotalpha_{\phantom{\dotalpha}\dotbeta} = 2g^{\mu\nu}\delta^\dotalpha_{\phantom{\dotalpha}\dotbeta}\,.
\end{equation}
This enables the construction of gamma matrices, which fulfil the Clifford algebra \cite{Dreiner:2008tw}.
The gravitational covariant derivative $\nabla_{\mu}$ contains no SM gauge fields but the spin connection. The spin connection is not an independent field, but depends on the metric through the vielbein. Accordingly, in O(4) symmetric gauge \cite{vanNieuwenhuizen:1981uf,Woodard:1984sj}, quantum fluctuations of the vielbein and the spin connection can be rewritten entirely in terms of metric fluctuations (see App.~\ref{app:weyl_fermions_in_qg}).
Spacetime indices are understood to be contracted with the full metric $g_{\mu\nu}$.

The beta functions of the four-fermion couplings have a gravitational and a non-gravitational contribution,
\begin{equation}
\beta_{c_i}= \beta_{c_i}^{\rm F}+ \beta_{c_i}^{\rm grav.}, \quad \forall c_i\in \set{c_{qq},c_{qu},c_{qd},c_{uu},c_{ud},c_{dd}, \cRe, \cIm}.
\end{equation}
The fermion contributions read
\begin{align}
	\label{eq:beta_cqq_without_gravity}
	\beta_{\cqq}^{\rm F}   & = 2 \cqq + l_{1}^{(F)} v_{4}\Big[4 \cqq^2 + \cqd^2 +  \cqu^2\Big] \,,\\
	\label{eq:beta_cqu_without_gravity}
	\beta_{\cqu}^{\rm F}   & = 2 \cqu + l_{1}^{(F)} v_{4}\Big[6 \cqu^2 + 12 \cqq \cqu + \frac{9}{2} {\cRe}^2 + \frac{9}{2} {\cIm}^2 + 8 \cqu \cuu + 2 \cqd \cud\Big] \,,\\
	\label{eq:beta_cqd_without_gravity}
	\beta_{\cqd}^{\rm F}   & = 2 \cqd + l_{1}^{(F)} v_{4}\Big[6 \cqd^2 + 12 \cqq \cqd + \frac{9}{2} {\cRe}^2 + \frac{9}{2} {\cIm}^2 + 8 \cqd \cdd + 2 \cqu \cud\Big]\,, \\
	\label{eq:beta_cuu_without_gravity}
	\beta_{\cuu}^{\rm F}   & = 2 \cuu + l_{1}^{(F)} v_{4}\Big[2 \cqu^2 + \cud^2\Big] \,,\\
	\label{eq:beta_cud_without_gravity}
	\beta_{\cud}^{\rm F}   & = 2 \cud + l_{1}^{(F)} v_{4}\Big[- 6 \cud^2 + 4 \cqd \cqu + 8 \cud \cdd  + 8 \cud \cuu\Big]\,,\\
	\label{eq:beta_cdd_without_gravity}
	\beta_{\cdd}^{\rm F}   & = 2 \cdd + l_{1}^{(F)} v_{4}\Big[2 \cqd^2 + \cud^2\Big]\,,\\
	\label{eq:beta_cquqd_without_gravity}
	\beta_{\cRe}^{\rm F} & = 2 \cRe + l_{1}^{(F)} v_{4}\Big[12 \cqu  + 12 \cqd  \Big]\cRe\,,\\
    \beta_{\cIm}^{\rm F} & = 2 \cIm + l_{1}^{(F)} v_{4}\Big[12 \cqu + 12 \cqd \Big] \cIm\,.
	\label{eq:beta_cquqd_complex_conjugated_without_gravity}
\end{align}
The threshold function $l_1^{(F)}$ can be set to one with an appropriate choice for the regulator (see Eq.~\eqref{eq:shape_function}) and $v_4 = 1/32\pi^2$.

The system of flow equations is invariant under an $\mathrm{O}(2)$ rotation of the couplings $(\cRe, \cIm)$ and these couplings can therefore be expressed as a modulus and an angle.\footnote{Alternatively, one may express this in the complex basis as an independence from the phase of $\cquqd$.}
The absolute value of the combination  $((c_{quqd}^{+})^2+(c_{quqd}^{-})^2)$
is fixed by Eqs.~\eqref{eq:beta_cqu_without_gravity}, \eqref{eq:beta_cqd_without_gravity}, but the angle is undetermined.
Without loss of generality, we fix the value of the angle such that $\cIm$ is always zero and obtain a system of 7 equations for $\{c_{qq},c_{qu},c_{qd},c_{uu},c_{ud},c_{dd},\cRe\}$. 
We provide an overview of the zeros of these beta functions, i.e., the possible fixed-point candidates, in App.~\ref{app:fixed_points}.
These can in principle be interesting in their own right, and result in deviations from canonical scaling behavior. This can in particular be interesting in the context of composite Higgs models \cite{Bellazzini:2014yua,Sanz:2017tco,Witzel:2019jbe}, but we do not analyze it further here.

The propagator of metric fluctuations is derived from the Einstein-Hilbert truncation
\begin{equation}
	\label{eq:EH_action}
	\Gamma_k^{\text{EH}} = \frac{1}{16\pi G_N} \int \dd[4]{x \sqrt{g}} (2\Lambda - R)\,,
\end{equation}
with the Newton coupling 
\begin{equation}
G_N = g / k^2,
\end{equation}
and cosmological constant 
\begin{equation}
\Lambda = \lambda k^2.
\end{equation}
We work in an off-shell setting for metric fluctuations, i.e., we keep a non-zero cosmological constant in the metric propagator about a flat background.
We keep the dimensionless Newton coupling and cosmological constant, $g$ and $\lambda$ as free parameters, rather than evaluating their fixed-point values. This allows us to explore the possible gravitational parameter space more broadly. In this way, we account for two distinct uncertainties: the first uncertainty is a purely calculational one and means that due to the necessity to truncate the full dynamics, the gravitational fixed-point values can only be calculated with systematic uncertainties. The second uncertainty is physical: the gravitational fixed-point values depend on the number of matter fields that are present. Thus, if, in addition to the SM, there is, e.g., an extended dark sector, the gravitational fixed-point values change. We aim for statements about the four-fermion interactions which are robust with respect to both of these uncertainties.

We find the following gravitational contributions to the beta functions
\begin{align}
    \label{eq:induced_beta_fct1}
    \beta_{c_i}^{\rm grav.} &=  I(\lambda) g^2 + c_i L(\lambda) g \quad \forall c_i\in \set{c_{qq},c_{uu},c_{dd}}, \\ 
    \label{eq:induced_beta_fct2}
    \beta_{c_i}^{\rm grav.} &= -2 I(\lambda) g^2 + c_i L(\lambda) g \quad \forall c_i\in \set{c_{qu},c_{qd}}, \\ 
    \label{eq:induced_beta_fct3}
    \beta_{c_{ud}}^{\rm grav.} &= 2 I(\lambda) g^2 + c_{ud} L(\lambda) g\,, \\ 
    \label{eq:non-induced_beta_fct}    
\beta_{\cRe}^{\rm grav.} &= \cRe\, L(\lambda) g\,.
\end{align}
Gravitational contributions are proportional to one power of the dimensionless Newton coupling $g$ for each graviton propagator that occurs in the corresponding diagram.\footnote{The counting of $g$ is of course equivalent, if metric fluctuations are rescaled by a power of $\sqrt{G_N}$, such that $\sqrt{G_N}^n$ occurs at each vertex with $n$ gravitons, resulting in the same counting of $g$ after rescaling to the dimensionless Newton coupling.}
Two distinct types of terms arise from gravity: The terms $\sim I(\lambda)g^2$ \emph{induce} four-fermion interactions. They prevent a fixed point at vanishing four-fermion coupling as soon as $g_{\ast}\neq 0$. In contrast, the terms $\sim L(\lambda) g$ change the effective scaling dimension of the four-fermion coupling. They do not prevent a fixed point at vanishing four-fermion coupling. In combination, the fixed-point values for $c_{qq}, c_{qu}, c_{qd}, c_{uu}, c_{ud}, c_{dd}$ cannot be zero, but that for $\cRe$ can.

The dependence on the dimensionless cosmological constant $\lambda = \Lambda k^{-2}$ is given by
\begin{equation}
    \label{eq:grav_constants}
    I(\lambda)= -\frac{5}{8(1-2\lambda)^3}\,,\quad
     L(\lambda)= \frac{5}{2 \pi (1-2\lambda)^2} - \frac{93}{20 \pi (3-4\lambda)^2} - \frac{3}{20 \pi (3-4\lambda)}\,.
\end{equation}
We can identify the contribution of the transverse traceless mode, which diverges for $\lambda = 1/2$, while the trace mode has a pole at $\lambda = 3/4$. 
The transverse traceless mode is unaffected by the choice of gauge condition, whereas other choices of covariant gauge conditions result in a change of the pole in the propagator of the trace mode. We can see that the trace mode contribution is subdominant compared to the transverse traceless contribution, thus we expect a subleading gauge dependence of our results.

Crucially, for $\beta_{c_{quqd}}$, the inducing term $I(\lambda)g^2$ is absent.
This is because the required bubble diagram \raisebox{-3.8mm}{\includegraphics[height=10mm]{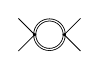}} cannot be constructed from minimal coupling of gravity to the kinetic terms of the fermions.
In fact, this is due to the way in which global symmetries determine the structure of the asymptotically safe fixed point, see Sec.~\ref{sec:globalsymmetries}. 
We come back to this point in more detail below.

\subsection{Fixed point in the maximally symmetric asymptotically safe theory space}
\label{sec:msas_toymodel}

The purely fermionic system, i.e., the system of beta functions with $g$ set to zero, admits $22$ fixed-point candidates with varying number of relevant directions (see Tab.~\ref{tab:fermionic_fps} in App.~\ref{app:fixed_points}). 
One of these is the fully irrelevant Gaussian fixed point.
Turning on gravitational fluctuations, there are also beta functions for $g$ and $\lambda$ which determine their fixed-point values $g_{\ast}$ and $\lambda_{\ast}$. As motivated above, we leave $g_{\ast}$ and $\lambda_{\ast}$ as free   parameters below.

When we turn on gravitational fluctuations by allowing $g_{\ast}$ to grow from zero, the fixed points for the four-fermion couplings move in theory space as functions of $g_*$ and $\lambda_*$. 
We track these movements, mostly focusing on the shifted Gaussian fixed point.
This is the unique fixed point that is maximally symmetric and continuously connected to the fully IR attractive Gaussian fixed point in the limit $g_* \to 0$.
Within the maximally symmetric theory space, perturbations are always fully irrelevant at the SGFP.

The maximally symmetric theory space in our case is one-dimensional, spanned by a single coupling $c_A$, which is related to the seven four-fermion couplings that we consider by
\begin{equation}
    c_{A}=-c_{qq}= -c_{uu}=-c_{dd}= -\frac{c_{ud}}{2}= \frac{c_{qu}}{2}= \frac{c_{qd}}{2}\,,\quad \cRe =0\,. 
    \label{eq:MSAS_relation_for_c}
\end{equation}
These relations arise because
the kinetic term in
Eq.~\eqref{eq:fermion_kinetic_action} is symmetric under a larger global symmetry group than $\mathrm{SU}(2)_L$. It is actually invariant under rotations of all the different Weyl fermions into each other.
This constitutes a $U(N_W)$  symmetry when considering $N_W$ Weyl fermions. There is a single four-fermion interaction compatible with $U(N_W)$ symmetry, namely
\begin{equation}
\Gamma_{k}^{4F,\, \rm MSAS} = \frac{c_A}{k^2}\big(\psi^{\dagger}_I\bar{\sigma}\,\psi^I\big)\big(\psi_J^{\dagger}\bar{\sigma}\,\psi^J\big)\,,
\label{eq:4F_action_msas}
\end{equation}
where $\psi^I =(q_a, u^c, d^c)$ is a vector of two-component spinors.
Each of the components denotes one of the Weyl spinors in our setting.
Eq.~\eqref{eq:4F_action_msas} in fact describes the axial current, see Section~\ref{sec:light_fermions_take_5}.

The beta function for $c_A$ is given by 
\begin{equation}
\beta_{c_A} = \big(2+ g\,L(\lambda)\big) c_A -g^2 \,I(\lambda) - \frac{3 c_A^2}{8\pi^2}\,.
\end{equation}
One can check explicitly that the beta functions in Eq.~\eqref{eq:induced_beta_fct1}-\eqref{eq:induced_beta_fct3} reduce to this expression with the identification in Eq.~\eqref{eq:MSAS_relation_for_c}. 
For this, it is crucial that the inducing term $I(\lambda)$ is absent in $\beta_{\cRe}$.
This allows us to track the shifted Gaussian fixed point efficiently, because we only need the solution to $\beta_{c_A}=0$, which is given by
\begin{equation}
c_{A\,\ast}= \frac{4\pi^2}{3}\left(2+ g\,L(\lambda)- \sqrt{\big(2+g\,L(\lambda)\big)^2- \frac{3}{2\pi^2}g^2\,I(\lambda}) \right),
\label{eq:sgfp_axial_coupling}
\end{equation}
The fixed point is characterized by a single critical exponent within the MSAS theory space,
\begin{equation}
\theta_{\rm MSAS}= -\sqrt{\big(2+ g\,L(\lambda)\big)^2 - \frac{3}{2\pi^2}g^2\,I(\lambda)}\,.\label{eq:thetaMSAS}
\end{equation}
Because $I(\lambda)<0$ for $g>0$ and $\lambda<1/2$, the fixed point is real everywhere. The critical exponent is increasingly negative, as $g$ increases.

The analysis of the SGFP so far guarantees its existence and shows that the critical exponent within the MSAS theory space  (i.e.\ along $c_A$) is always negative. 
Therefore, the direction along the line that satisfies Eq.~\eqref{eq:MSAS_relation_for_c}, is irrelevant and results in a prediction for the coupling.
There are, however, 6 additional critical exponents that characterize the fixed point. They are associated to a set of nested subspaces of the theory space, with increasingly lower degrees of symmetry. For the largest theory space, which includes $\cRe$, this symmetry is $\mathrm{SU}(2)_L\times\mathrm{U}(1)_Y$. Intermediate symmetries and corresponding constraints on couplings are listed in App.~\ref{app:fixed_points}.

These additional critical exponents have not been explored in previous work; thus it is not guaranteed that all eigenperturbations of the shifted Gaussian fixed point are irrelevant. If one of them is relevant, then EFT ordering is no longer the appropriate ordering for SMEFT coefficients in asymptotic safety.

\subsection{Stability trading during fixed-point collisions}
\label{sec:FP_collisions}

The IR value of a coupling is not automatically zero or non-zero just because its fixed-point value is so. If the coupling is relevant at the fixed point, virtually any IR value may be achieved (subject to global constraints on the RG flow), irrespective of the UV value. In particular, a coupling that is relevant at the fixed point may circumvent naturalness expectations, if it is canonically irrelevant. Contrary to the canonical expectation, such a coupling may have a Planck-scale value significantly larger than 1, even if its fixed-point value is $\mathcal{O}(1)$. This translates into a corresponding, ``unnaturally'' large value for the SMEFT coefficient in the IR.\footnote{In a converse application of the same idea, it has been proposed in \cite{Wetterich:2016uxm} that the Higgs mass parameter may become irrelevant in asymptotic safety, which would provide an ``unnaturally'' small value at the Planck scale and automatically give rise to a large hierarchy between Planck scale and electroweak scale.} 

Against this background, determining which SMEFT coefficients have non-vanishing fixed-point values only provides us with a part of the information that we need to  determine whether or not the SMEFT coefficients follow the EFT ordering that is usually assumed for them.

Many studies in asymptotic safety conclude near perturbative behavior of gravity for the UV-fixed point \cite{Falls:2013bv,Falls:2014tra,Falls:2017lst,Falls:2018ylp,Kluth:2020bdv,Becker:2024tuw,Eichhorn:2018ydy,Eichhorn:2018akn,Eichhorn:2018nda, Assant:2026dca}. This means that the critical exponents follow the canonical dimension of couplings relatively closely \cite{Eichhorn:2011pc,Eichhorn:2016esv,Eichhorn:2017eht,Christiansen:2017gtg,Eichhorn:2021qet,Eichhorn:2020sbo,Laporte:2021kyp,deBrito:2021pyi,deBrito:2021akp,deBrito:2025ges,Assant:2025gto,deBrito:2023myf}, see also the discussion in the review \cite{Eichhorn:2026uqj}. For matter couplings, this implies a gravitational anomalous dimension that shifts the scaling dimension of a coupling away from the canonical one by a relatively small amount. 
This leads us to the expectation that dimension-six-interactions should stay irrelevant, such that the dimensionless Wilson coefficients at the Planck scale are given by $\mathcal{O}(1)$ fixed-point values and there is no mechanism to drive these coefficients up to huge values which would (partially) counteract Planck-scale suppression for them.

More specifically, the shifted Gaussian fixed point has fixed-point values of order $\mathcal{O}(g_*)$, such that the critical exponents of the four-fermion couplings have deviations from $-2$ which are parametrically controlled by $g_{\ast}$.

\begin{figure}[!ht]
    \centering
	\begin{subfigure}[b]{.4\textwidth}
		\centering
		\includegraphics[width=\textwidth]{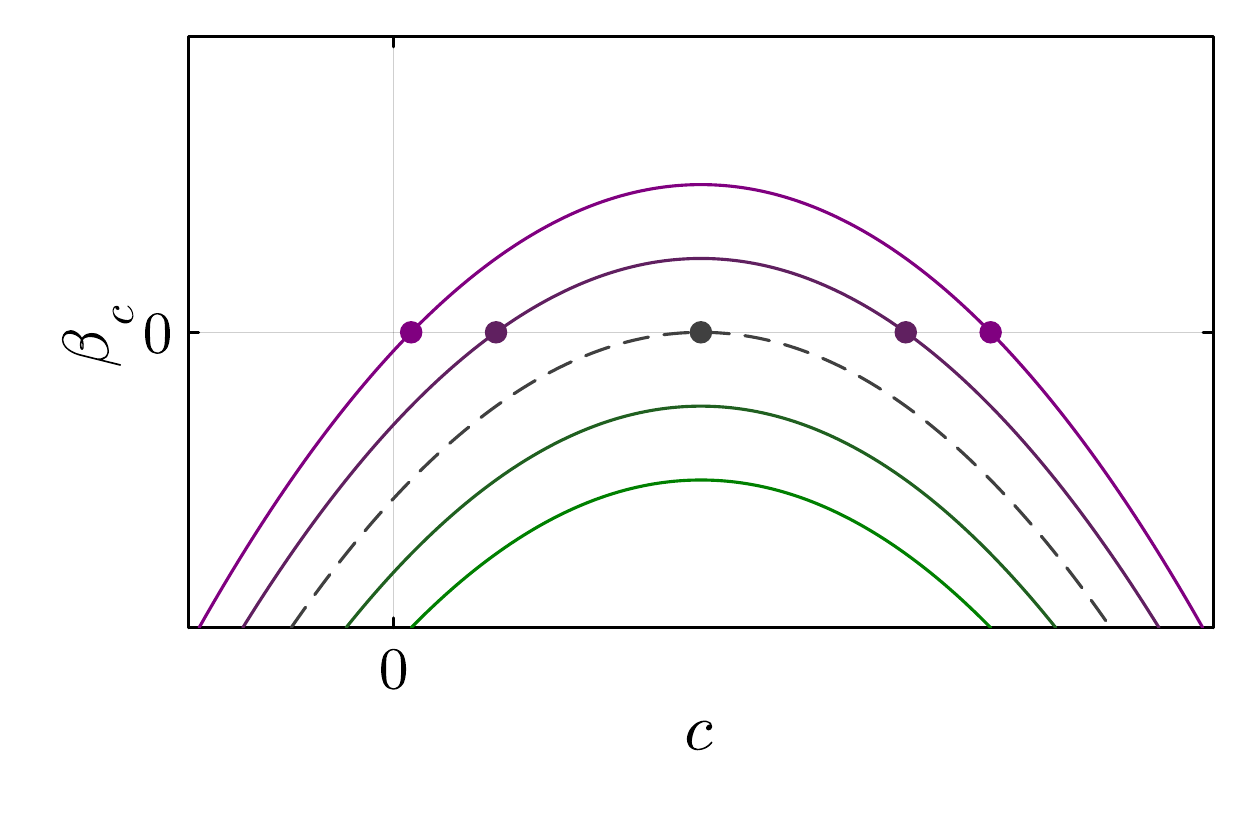}
		\caption{\centering Fixed points annihilate for induced couplings for which
        $b\neq0$
        .}
		\label{fig:fp_annihilation}
	\end{subfigure}
	\begin{subfigure}[b]{.4\textwidth}
		\centering
		\includegraphics[width=\textwidth]{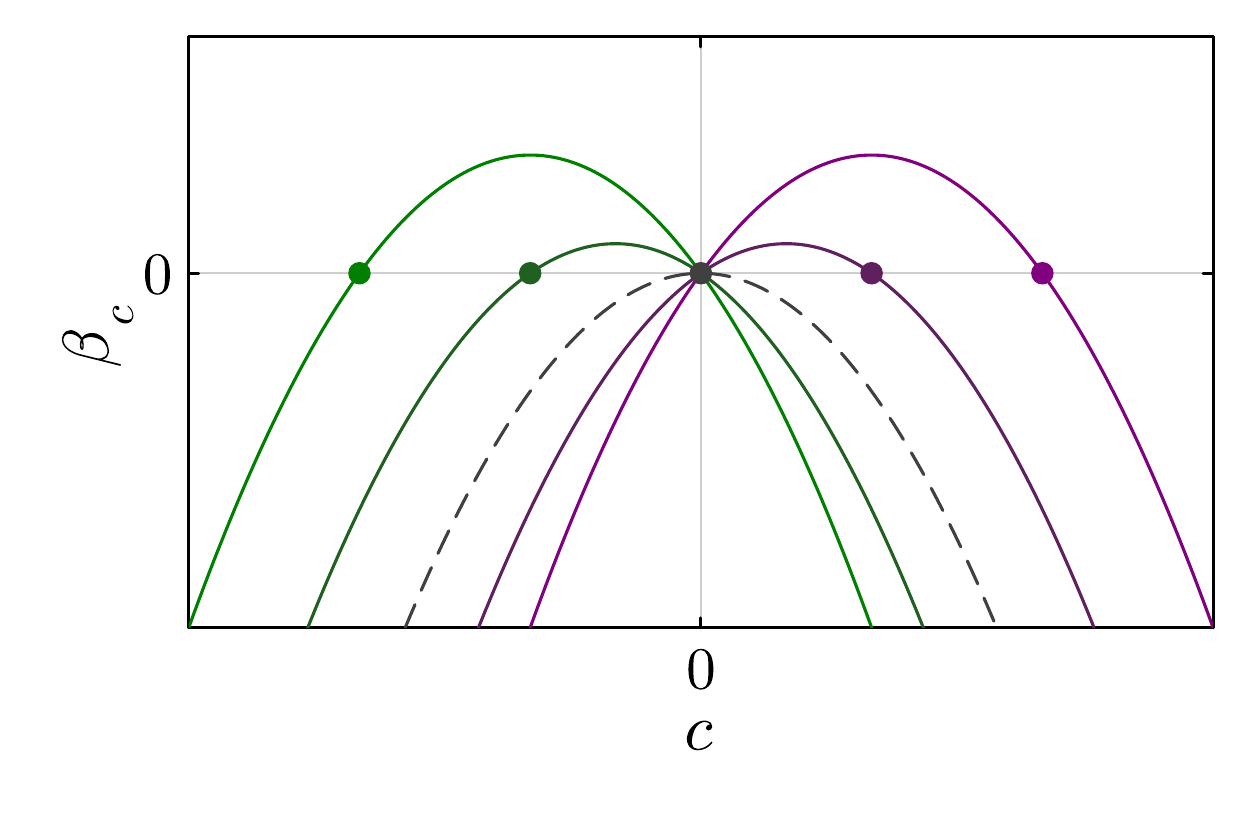}
		\caption{\centering Fixed point stability trading can occur for couplings with
        $b=0$.
        }
		\label{fig:fp_stability_trading}
	\end{subfigure}
    \caption{The result of fixed-point collisions depends on the structure of the flow. 
        For a simplified quadratic flow $\beta_c = \beta_c^{\text{flat}} + a g c + b g^2$ depending on the gravitational coupling $g$ with some coefficients $a$ and $b$ we plot example fixed points with couplings above, below and at the critical value of $g$. 
        At the critical value, marked with a dashed line, the fixed point becomes degenerate in both cases.
        In \ref{fig:fp_annihilation}, the fixed points collide and annihilate. 
        In \ref{fig:fp_stability_trading}, both fixed points exist before and after the collision, but the sign of the critical exponent changed, which in this one-dimensional example is simply a change in the slope.}
	\label{fig:fp_collisions}
\end{figure}

Thus, a change of a four-fermion interaction from canonical irrelevance to a relevant interaction requires that the real part of a critical exponent goes through zero as a function of $g_{\ast}$. 
There are two mechanisms for this to occur.
The first mechanism can be realized if a critical exponent is complex.
This enables its real part to switch sign as a function of $g_{\ast}$, see \cite{deBrito:2023myf} for an example.
The second mechanism is more common and involves at least two fixed points.
If the critical exponent of a fixed point is real, it can only switch sign if the fixed point is degenerate at this point, i.e., if two (or more) fixed points approach each other as a function of $g_{\ast}$ and collide at a critical value of $g_{\ast}$.
However, such a collision can happen in two distinct ways \cite{deBrito:2023kow}, only one of which provides a relevant direction:
\begin{description}
    \item[Scenario I:] The fixed points annihilate, as for values of gravitational couplings beyond the boundary, the zeros of the beta functions become complex, see \cite{Eichhorn:2017eht,Christiansen:2017gtg, deBrito:2021pyi, Eichhorn:2021qet} for examples.\footnote{These examples work with truncations of the RG flow and fixed-point collisions occur close to the boundary of the regime in which the truncation is reliable, see \cite{deBrito:2023myf} for a cautionary example.}
        As a consequence, the point where the critical exponent is zero constitutes a barrier for the existence of the fixed point and there is no actual change in the number of relevant directions of the physically viable (real) fixed point.
    \item[Scenario II:] The fixed points exchange stability properties, i.e., the critical exponent of one of the fixed points becomes positive, while the other becomes negative. Such stability trading is relevant to many examples in condensed-matter systems \cite{Calabrese:2002bm,Eichhorn:2013zza,Herbut:2015zqa,Ray:2023nla}.
\end{description}
These scenarios are displayed in Fig.~\ref{fig:fp_collisions}.

In collisions of more than two fixed points, a combination of these scenarios may take place.

\begin{figure}[!ht]
	\begin{center}
	\includegraphics[width=0.7\textwidth]{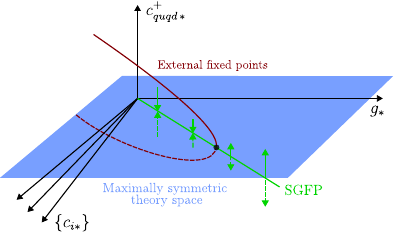}
    \end{center}
	\caption
    {
        Schematic illustration of the fixed-point collision between the shifted Gaussian fixed point and the fixed points external to the maximally symmetric subspace observed in our toy model.
        The maximally symmetric theory space is parametrized by the $\{c_i\}$.
            The SGFP assumes some value inside of this space as a function of $g_*$.
            The pair of external fixed points approach from outside the maximally symmetric subspace.
        Arrows indicate the irrelevance of the SGFP for low $g_*$ and relevance beyond the collision.
Since we fixed $\cIm = 0$, there is a pair of solutions with relative sign in $c_{quqd\,*}^+$ which represent infinitely many fixed points converging on that collision point.
}   
	\label{fig:num_fps}
\end{figure}

Whether either of the two scenarios or only scenario II can be realized, is determined by the symmetry properties of the space of couplings under consideration. 
Symmetries define closed subspaces of the space of couplings. The RG flow cannot leave a given subspace.
Therefore, fixed-point collisions that occur between fixed points \emph{within} the same subspace can in principle realize scenario I or II. In contrast, fixed-point collisions that occur between fixed points in distinct subspaces (of which one must contain the other), can only occur if the fixed point of lower symmetry realizes the higher symmetry at the collision point, i.e., if it traverses the higher-symmetry-subspace. During its traverse, it can encounter the fixed point that lives in the higher-symmetry subspace.
Such a fixed-point collision can only realize scenario II. 
This is because one of the fixed points participating in the collision lies within a subspace of higher symmetry. It can therefore be tracked as a function of $g_{\ast}$ without any knowledge of the RG flow within the larger subspace. 
Accordingly, it cannot become complex as a consequence of a collision with a fixed point that moves within the larger subspace.
One might say that this fixed point is protected by its enhanced symmetry.
At such a collision, only its critical exponents associated to directions within the larger subspace, orthogonal to the smaller subspace, can change, see Fig.~\ref{fig:num_fps}.

For the SM fields, the maximally symmetric theory space, spanned by all couplings compatible with the maximal global symmetries of the kinetic operators, contains all those interactions that have non-zero values in the asymptotically safe UV regime.
Technically, this is realized by induction terms in the beta function (which are independent of the coupling itself), as in Eq.~\eqref{eq:induced_beta_fct1}-\eqref{eq:induced_beta_fct3}.
For the four-fermion couplings, all except $\cRe$ lie in this maximally symmetric subspace, but have to satisfy Eq.~\eqref{eq:MSAS_relation_for_c} to lie within this space, i.e., they are not all independent. The shifted Gaussian fixed point, which becomes the Gaussian fixed point for $g_{\ast} \rightarrow 0$, lies in this subspace.\footnote{It can in fact be mapped to the shifted Gaussian fixed point in \cite{Eichhorn:2011pc} by the map $c_A = \frac{1}{2}\left(\lambda_{+}-\lambda_{-} \right)$.}

The existence of the sGFP for all values of $g_{\ast}$ and $\lambda_{\ast}$ that are physically viable is well-established, see Sec.~\ref{sec:action}. In other words, it is known that scenario I cannot occur for the shifted Gaussian fixed point.

Couplings outside the MSAS theory space vanish at the shifted Gaussian fixed point. 
Thus, $c_{quqd\,*}^{+}=0$ and additionally all deviations from Eq.~\eqref{eq:MSAS_relation_for_c} vanish.
However, there are further fixed points in the larger theory space, which do not satisfy  Eq.~\eqref{eq:MSAS_relation_for_c} and feature $c_{quqd\,*}^{+}\neq 0$. 
These additional fixed points have not been studied previously. In themselves, we do not consider them suitable choices for UV fixed points, because they come with at least one relevant direction already at $g_{\ast}=0$, i.e., they decrease the predictivity of the fixed point. They can, however, collide with the sGFP and undergo the stability trading scenario II. 

\begin{figure}[!t]
\begin{center}
\includegraphics[width=0.5\linewidth]{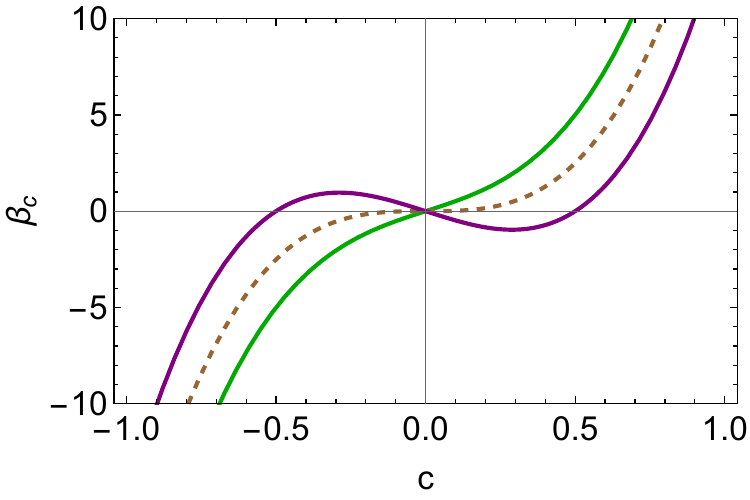}
\end{center}
\caption{\label{fig:illustration_collision}We show a beta function for a coupling $c$ which illustrates the type of fixed-point collision we investigate: One fixed point always lies at $c_{\ast}=0$. The other fixed points are initially (purple curve) real and located on both sides of $c_{\ast}=0$. They approach each other and eventually form a triply degenerate fixed point (brown dashed curve). Beyond the collision, only the fixed point at $c_{\ast}=0$ remains; the other two have become complex.}
\end{figure}

Indeed, we find the following situation to be realized, see Fig.~\ref{fig:num_fps}: there are two fixed points with non-zero $c_{quqd\,\ast}^{+}$, which feature opposite sign of $c_{quqd\,\ast}^{+}$. For large $g_{\ast}$, $c_{quqd\,\ast}^{+}$ decreases for both fixed points with $g_\ast$.
Therefore, there is a fixed-point collision between these two fixed points at $c_{quqd\,\ast}^{+}=0$ at a critical value $g_{\ast\, \rm crit.}(\lambda_{\ast})$. At this fixed-point collision, the fixed-point values of all other couplings satisfy Eq.~\eqref{eq:MSAS_relation_for_c}.
Therefore, at the fixed-point collision, the shifted Gaussian fixed point is a triply degenerate fixed point. For $g_{\ast}>g_{\ast\,\rm crit.}$, only one of the three real fixed points remains, namely the shifted Gaussian fixed point. The other two fixed points become complex. The situation is illustrated in Fig.~\ref{fig:illustration_collision}, which is a combination of the two scenarios in Fig.~\ref{fig:fp_collisions}.

This collision with the two lower-symmetry fixed points transfers a relevant direction orthogonal to the MSAS subspace to the SGFP.
The resulting relevant interaction is the non-induced interaction of lesser symmetry. 
The fixed-point value for this coupling remains zero, but starting from this fixed-point value, it may become non-zero in the IR.

\begin{figure}[!ht]
	\begin{center}
    \begin{subfigure}[b]{.491\textwidth}
        \includegraphics[width=1.0\textwidth]{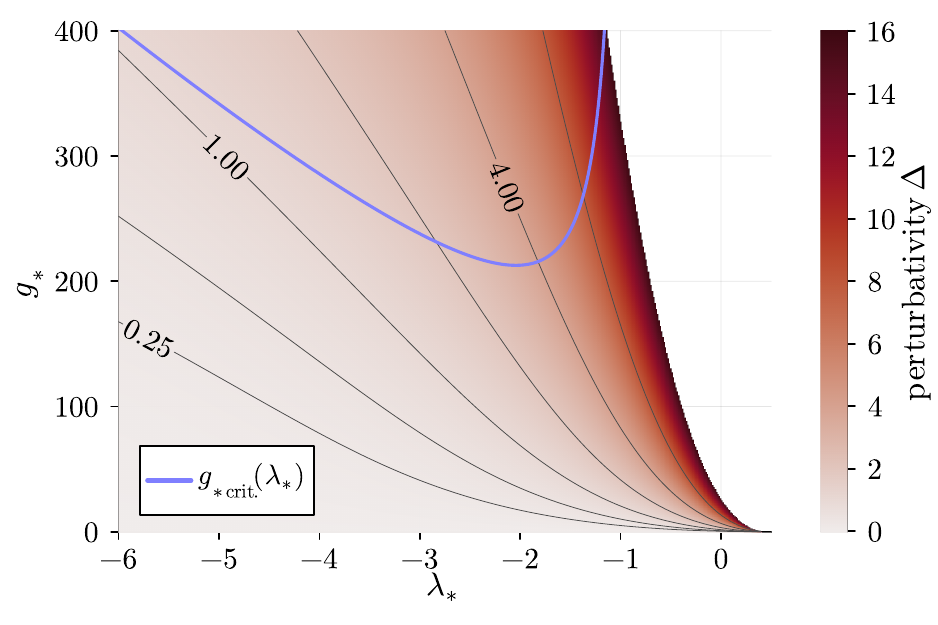}
    \end{subfigure}
    \hfill
	\begin{subfigure}[b]{.491\textwidth}
        \includegraphics[width=1.0\textwidth]{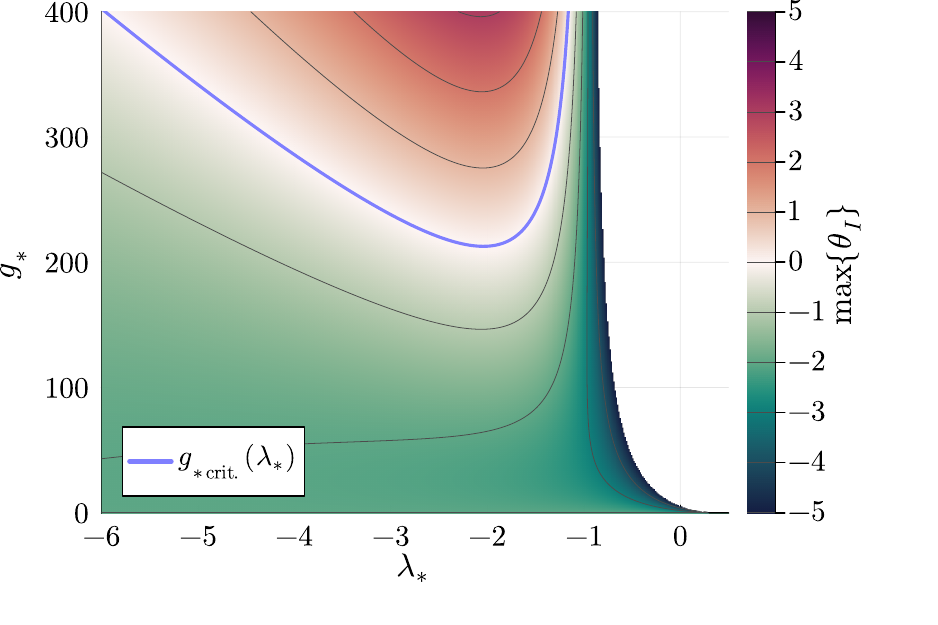}
    \end{subfigure}
    \end{center}
	\caption
    {Perturbativity and maximum of the set of critical exponents of the shifted Gaussian fixed point.
    Perturbativity is measured by the quadratic deviation from the canonical scaling $\Delta$ and capped to a maximum of $16$. 
    We provide extended plots in App.~\ref{app:extended_plots}. }
	\label{fig:perturbativity}
\end{figure}

We evaluate the value $g_{\ast\,\rm crit.}(\lambda)$ (see App.~\ref{app:extended_plots}), cf.~right panel in Fig.~\ref{fig:perturbativity}. We find a relatively high value, compared to ``typical'' fixed-point values in the literature. 
One can understand the large values of $g_*$ for the transition $g_{\ast\,\rm crit}(\lambda)$ to stem from the ratio of loop- to canonical scaling effects, where loop effects are generically suppressed by $1/16\pi^2$ \cite{Eichhorn:2017eht}. 
To lift this suppression, gravity must be correspondingly non-perturbative.
This makes it somewhat unlikely that the scenario of a fixed-point collision is realized.\\
Besides the question, whether ``typical'' fixed-point values fall into the vicinity of $g_{\ast\,\rm crit.}$, it is important to check whether our truncation is trustworthy at these values. We do so in two ways: First, we calculate the expected deviation from canonical scaling 
\begin{equation}
\Delta = \sqrt{\frac{1}{7} \sum_{i=1}^{7} \left(\theta_i - d_{\bar{c}_i}\right)^2}\,,
\label{eq:def_perturbativity}
\end{equation}
where $\theta_i$ are the critical exponents and $d_{\bar{c}_i}$ is the canonical dimension of the couplings; in our case $d_{\bar{c}_i}=-2$.
This measure was introduced for gravity-matter systems in \cite{Eichhorn:2020sbo}, following similar ideas for pure gravity in \cite{Falls:2013bv, Falls:2014tra, Falls:2017lst, Falls:2018ylp}, to calculate the typical size of the gravity-induced anomalous scaling dimension for matter interactions across the $(g_{\ast},\lambda_{\ast})$-plane.
We find that the curve $g_{\ast\,\rm crit. }(\lambda)$ intersects contours of constant $\Delta$, see left panel of Fig.~\ref{fig:perturbativity}.
We find that, despite the relatively large value of $g_{\ast}$, we can remain close to a near-perturbative regime, where the deviation from canonical scaling is not larger than one.
We consider this as a sign of the reliability of our results and in particular of the fixed-point collision that it occurs in a regime where we have likely accounted for all relevant fermionic interactions.
All terms beyond our truncation are either higher than dimension six or exactly dimension six but for non-minimal fermion-curvature couplings \cite{Eichhorn:2016vvy, Daas:2020dyo, Daas:2021abx}.
None of these is likely to be a relevant interaction where the fixed-point collision occurs, if we assume that the anomalous dimension for all interactions is approximately the same.

With an outlook to generalizing beyond four-fermion couplings towards all interactions in the SMEFT, we summarize as follows.
The impact of gravity on higher order matter operators can thus be partitioned into two classes.
Those that respect maximal symmetry, i.e., the symmetries of the kinetic operators -- these couplings are part of the MSAS subspace -- and those that do not -- they lie outside the MSAS subspace.
The former can impose constraints on the gravitational theory space (i.e., weak-gravity bounds \cite{Eichhorn:2016esv, Eichhorn:2017eht, Christiansen:2017gtg, deBrito:2021pyi, Eichhorn:2021qet, deBrito:2023myf}), but generically remain irrelevant and are therefore Planck-scale suppressed in the IR.\footnote{It is not ruled out that stability trading occurs for induced operators. But it would require the beta function is zero, together with a critical exponent and its derivative, i.e., a saddle point in the beta function at finite coupling.
The simple picture of an induced operator may change upon inclusion of higher orders, as seen in \cite{deBrito:2023myf}.} In contrast, the latter are zero at the SGFP, but can become relevant due to fixed-point collision. Therefore, those couplings outside the MSAS theory space are the ones that can circumvent Planck-scale suppression.

As an upshot, the most predictive fixed point in asymptotic safety generically renders all four-fermion couplings that we have studied irrelevant. Accordingly, starting from the $\mathcal{O}(1)$ fixed-point values for these couplings, we arrive at a $\mathcal{O}(M_{\rm Planck}^{-2})$ suppression of the corresponding SMEFT coefficients in the IR. The $\mathcal{O}(M_{\rm Planck}^{-2})$ suppression follows, because the RG flow below the Planck scale is dominated by the canonical scaling term.
For couplings outside the MSAS the suppression is even stronger in the sense that the predicted IR value is zero. 

There is, however, the possibility that couplings that lie outside the MSAS theory space become relevant. This can occur after stability trading during a fixed-point collision of the shifted Gaussian fixed point.
Then, such a coupling retains its vanishing fixed-point value at the shifted Gaussian fixed point, but can achieve a large Planck-scale value.
As a consequence, it can, at least partially, circumvent the Planck-scale suppression.

\section{Light fermions in quantum gravity: Take five}
\label{sec:light_fermions_take_5}

The relative lightness -- compared to the Planck scale -- of fermions constitutes a promising testing ground for candidate quantum gravity theories. One may expect that gravity, already being a universally attractive force at the classical level, favors bound-state-formation linked to mass generation. In fact, another self-interacting gauge theory, namely QCD, provides a blueprint here, because it breaks chiral symmetry and generates massive bound states\cite{Braun:2011pp,Braun:2014ata}.
This argument motivated the first study of light fermions in asymptotically safe gravity \cite{Eichhorn:2011pc}, where quantum-gravity fluctuations were not found to favor bound-state formation. The diagnostic in this, and the subsequent works \cite{Eichhorn:2011pc, Meibohm:2016mkp, Eichhorn:2017eht, deBrito:2020dta, deBrito:2023kow}, see also \cite{deBrito:2023pli} for related work in quadratic gravity, were four-fermion couplings. They constitute tracers of dynamical mass generation. 
Four-fermion operators can  indicate the onset of a
condensation mechanisms of fermions which can be accompanied by the formation of massive bound states and generically leads to mass terms for these bound states.

Schematically, a fermion bilinear $(\bar{\psi}\psi) =: \phi$ acts as a scalar.
Partially bosonizing the action (using a Hubbard--Stratonovich transformation \cite{Hubbard:1959ub,stratonovich:1957}) in this way, one can identify
\begin{equation}
\lambda\, (\bar{\psi}\psi)^2 \rightarrow \frac{1}{\lambda}\phi^2 + y \phi \bar{\psi}\psi\,.
\end{equation}
The formation of a vacuum condensate is linked to a vanishing scalar mass, because this separates the regime of unbroken symmetry (with $\langle \phi \rangle=0$) from spontaneously broken symmetry (with $\langle \phi \rangle\neq 0$). 
Thus, a divergence in the four-fermion coupling signals the onset of condensate formation. The resulting spectrum of masses and bound states depends on the details of the four-fermion coupling(s) that diverge(s). 
In general, a divergence is, however, not compatible with the original fermions in the theory remaining light, unbound and propagating degrees of freedom.
For details in QCD, see \cite{Braun:2011pp,Braun:2014ata}, for similar considerations in lower-dimensional settings in condensed matter, see, e.g., \cite{Kopietz:2010zz,Metzner:2011cw}.

 Motivated by their use as a diagnostic for dynamical mass formation, four-fermion couplings were among the first self-interactions of matter fields to be studied in asymptotic safety \cite{Eichhorn:2011pc}.
 The general idea behind this study is that if quantum gravity favors divergent four-fermion couplings, the resulting mass scale should -- in the absence of very severe fine-tuning -- be of the order of the Planck scale.
 Even without knowing further details about condensation channels and further properties of the resulting bound states, it is clear that such a scenario would essentially be irreconcilable with the observation of light (compared to the Planck scale) fermions in the Standard Model.
 Following the first evidence \emph{against} divergent four-fermion couplings in asymptotic safety \cite{Eichhorn:2011pc}, see also \cite{Meibohm:2015twa}, in a simplified setting without gauge or Yukawa interactions and with global $\mathrm{SU}(N_f)_L \times \mathrm{SU}(N_f)_R$ symmetry (with $N_f$ denoting the number of flavors of Dirac fermions), further interactions were later added: In \cite{Eichhorn:2017eht}, the effects of a Yukawa coupling were added, again finding that quantum-gravity fluctuations do not induce critical behavior in four-fermion couplings.
 In \cite{deBrito:2020dta, deBrito:2023kow}, gauge interactions were added, confirming that they can trigger the onset of critical behavior, while the asymptotically safe system of gravity, gauge interactions and fermions can avoid this. \\
Simultaneously, the absence of critical behavior in fermionic systems has been used to constrain other aspects of asymptotic safety, linked to criticality induced by nontrivial gravitational backgrounds (rather than gravitational fluctuations) \cite{Gies:2018jnv, Gies:2021upb}.\\
In parallel, it was investigated that, if the gauge symmetries of the system allow the explicit addition of fermion mass terms, both Dirac and Majorana, then these mass terms generically have vanishing fixed-point values and are relevant \cite{Eichhorn:2016vvy, Daas:2020dyo, Daas:2021abx, DeBrito:2019rrh, deBrito:2025ges}. Thus, their IR value is a free parameter. For the SM fermions, this is of course not applicable, because the mass terms break the gauge symmetries, with the exception of Majorana mass terms for neutrinos, studied in asymptotic safety in \cite{DeBrito:2019rrh, deBrito:2025ges}.

The above series of studies, \cite{Eichhorn:2011pc, 
Eichhorn:2017eht, deBrito:2020dta, deBrito:2023kow}, while covering other aspects needed of a realistic description of SM fermions, neglected the Weyl nature of SM fermions. The present, fifth take on the question of fermionic masses, closes this gap.

From our discussion in the preceding sections, it has already become clear that four-fermion couplings feature a fixed point -- the shifted Gaussian fixed point -- at which they remain irrelevant for a very large interval in $g$. Accordingly, we do not find any evidence for critical behavior in four-fermion interactions. The only possible exception to this is in a channel that is only present due to the Weyl nature of SM fermions, namely $\cRe$, which can become relevant and therefore increasingly large, for very large $g$. The critical line beyond which this happens is shown in Fig.~\ref{fig:perturbativity}.
We conclude that, within the range of $g$ and $\Lambda$ that we consider credible for the gravitational fixed-point values, the system of four-fermion couplings stays far from criticality and bound-state-formation or condensation of bilinears is safely avoided.

In addition to investigating the fixed-point structure of the full set of four-fermion couplings, we may also consider a reduced set of degrees of freedom. This may be of interest, e.g., in modeling fermionic dark sectors
Three possibilities exist: \\
First, a system with just the left-handed doublet and correspondingly a single four-fermion interaction $c_{qq}$. It has two fixed points
\begin{equation}
    c_{qq\,*} = -4\pi^2 \left(2 + g L(\lambda) \pm \sqrt{\big(2+g L(\lambda)\big)^2 - \frac{g^2 I(\lambda)}{2\pi^2}}\right),
\end{equation}
using $L(\lambda)$ and $I(\lambda)$ defined in Eq.~\eqref{eq:grav_constants}.

These fixed points are real-valued for all positive $g$ and $\lambda<1/2$. One of them is the shifted Gaussian fixed point with negative critical exponent.\\
Second, one may consider a system with just the right-handed fermions and corresponding three interactions $c_{uu}$, $c_{dd}$ and $c_{ud}$. In that system, we find a shifted Gaussian fixed point at real values and negative critical exponents for all values of $g$ and $\lambda$ within a credible range for the gravitational fixed point.\\
Third, we may reduce the degrees of freedom to a single right-handed field, for which there is only the coupling $c_{uu}$, which has a beta function linear in $c_{uu}$. Accordingly, there is always a real-valued fixed point with negative critical exponent.

We conclude that quantum gravity fluctuations avoid the formation of fermionic condensates in all analyzed settings. 
\bigskip

To close the circle, let us come back to the results in \cite{Eichhorn:2011pc}, in which one may note that the gravitational contribution that generates non-zero fixed-point values for the two four-fermion couplings $\lambda_{\pm}$ has exactly opposite sign.
This means that of the two four-fermion interactions $\lambda_V\,(\bar{\Psi}^i\gamma^{\mu}\Psi^i)\,(\bar{\Psi}^j\gamma_{\mu}\Psi^j)$ and $\lambda_A\,(\bar{\Psi}^i\gamma^{\mu}\gamma_5\Psi^i)\,(\bar{\Psi}^j\gamma_{\mu}\gamma_5\Psi^j)$, only $\lambda_A$ is induced, although both $\lambda_V$ and $\lambda_A$ are required to form a Fierz-complete basis of $SU(N_F)_L \times SU(N_F)_R$ symmetric four-fermion interactions.
This suggests that there is in fact a larger symmetry in the kinetic term that is not manifest when the kinetic term is written in terms of Dirac fermions as $\bar{\Psi}^i\slashed{\nabla}\Psi^i$.
To see this, let us consider the following argument.

Weyl fermion bilinears can be formulated either in terms of right-handed -- e.g., $u^{\dot\alpha}$ -- or in terms of their left-handed (hermitian conjugate) representation $(u^c)_\alpha =\levicivita_{\alpha\gamma}u^{\dagger\,\gamma} $ (or vice versa; see \cite{Dreiner:2008tw}), such that
\begin{equation}
\begin{aligned}
    u^{c\,\dagger} \bar{\sigma}^\mu u^c &= 
    (u^{c\dagger})_{\dot\alpha}
    \bar{\sigma}^{\mu\,\dot\alpha\beta} (u^c)_\beta 
    = (\levicivita_{\alpha\gamma}u^{\dagger \gamma})^\dagger \bar{\sigma}^{\mu\,\dot\alpha\beta} (\levicivita_{\beta\delta}u^{\dagger \delta}) 
    = (u)^{\dot\gamma} \sigma^{\mu}_{\delta\dot\gamma} (u)^{\dagger\,\delta}
    = -u^\dagger \sigma^{\mu} u\,.
    \label{eq:kinetic_action_left_right}
\end{aligned}
\end{equation}

Using identity \eqref{eq:kinetic_action_left_right} and integrating by parts, the kinetic term can be written either in terms of $N_W/2$ Dirac fermions $\Psi_U$ and $\Psi_D$ or in terms of a vector of $N_W$ left-handed Weyl fermions $\psi^I$ (where $I \in \{1,\dots,N_W\}$), where
\begin{equation}
    \psi^I = \begin{pmatrix} q_1 \\ q_2 \\ u^c \\ d^c\end{pmatrix}^I,\quad
    \Psi_U = \begin{pmatrix} q_1 \\ u \end{pmatrix},\quad
    \Psi_D = \begin{pmatrix} q_2 \\ d \end{pmatrix}.
\end{equation}
Up to boundary terms, the kinetic part of the action reads
\begin{equation}
\begin{aligned}
    \Gamma_{\text{kin.}}^{D=4} &= \int\mathrm{d}^4{x}\;\Big\{q_a^\dagger \barsigma^\mu \nabla_\mu q_a + u^\dagger \sigma^\mu \nabla_\mu u + d^\dagger \sigma^\mu \nabla_\mu d \Big\} \\
    &= \int\mathrm{d}^4{x}\;\Big\{\psi^\dagger_I \barsigma^\mu \nabla_\mu \psi^I \Big\}\\
&= \int\mathrm{d}^4{x}\;\Big\{\bar{\Psi}_U \gamma^\mu \nabla_\mu \Psi_U + \bar{\Psi}_D \gamma^\mu \nabla_\mu \Psi_D \Big\}\,.
\label{eq:kinetic_dirac_weyl}
\end{aligned}
\end{equation}

The unification to Dirac fermions only works in case of uncharged fermions as in \cite{Eichhorn:2011pc}; a symmetry broken regime, where the Dirac fermion has a consistent charge; or at the MSAS, where the gauge couplings are expected to vanish.
We emphasize that while the formulation in terms of Dirac fermions is possible, it does not make the full $\mathrm{U}(N_W)$ symmetry of the system manifest.

Interpreting the fields in such a $\psi^I$ as left- and right-handed Weyl fermions, the bilinear of right-handed fermions acquires a relative minus with respect to the left-handed case from anticommuting according to Eq.~\eqref{eq:kinetic_action_left_right}.
For the case of $N_f =N_W/2$ uncharged Dirac fermions as investigated in \cite{Eichhorn:2011pc}, the kinetic term that determines the maximal symmetry of that system is equivalent to that in Eq.~\eqref{eq:kinetic_dirac_weyl}.
The construction of the maximally symmetric subspace applies and the four-fermion operator at the shifted Gaussian fixed point must look like Eq.~\eqref{eq:4F_action_msas}.
The SGFP of that system must then be aligned with the axial current.
In fact,
the maximally symmetric four-fermion current for our toy model
\begin{equation}
\begin{aligned}
    \mathcal{L}^{D=6} &= \big(\psi_I^\dagger \barsigma^\mu \psi^I\big)\big(\psi_J^\dagger \barsigma^\mu \psi^J\big) \\
    &= \big(q_a^\dagger \barsigma^\mu q^a - u^{c\,\dagger} \sigma^\mu u^c - d^{c\,\dagger} \sigma^\mu d^c\big)^2 \\
    &= \Big(\bar{\Psi}_U \gamma^\mu (P_L-P_R) \Psi_U + \bar{\Psi}_D \gamma_\mu(P_L-P_R)\Psi_D\Big)^2 \\
		  &= \Big(\bar{\Psi}_U \gamma^\mu \gamma_5 \Psi_U + \bar{\Psi}_D \gamma^\mu \gamma_5 \Psi_D\Big)^2 \\
\end{aligned}
\end{equation}
is purely axial.

It is, in fact, reassuring that gravity induces interactions of Weyl fermions, rather than the (less symmetric) four-fermion interactions of Dirac fermions. This fits with the fact that, coming out of the transplanckian regime with quantum gravity, the SM is based on Weyl fermions and not Dirac fermions as its fundamental building blocks. It raises the interesting question whether the chiral nature of fermionic matter leaves an imprint in gravity. For instance, it is conceivable that the quantum fluctuations of minimally coupled Weyl fermions induce parity-breaking in higher-order curvature operators. We leave this very interesting question for future work.

The formation of fermionic bound states, resulting from a strong-coupling regime, is also of interest in the context of effective asymptotic safety. Effective asymptotic safety \cite{Percacci:2010af, deAlwis:2019aud} is a scenario in which the asymptotically safe fixed point is never realized exactly in the deep UV. Instead, it is only realized approximately. There is a transplanckian UV cutoff scale $\Lambda_{\rm QFT,\, UV}$, beyond which the quantum field theoretic description of gravity and matter breaks down and, e.g., string theory takes over. At $\Lambda_{\rm { QFT,\, }UV}$, the couplings in the matter-gravity QFT are determined by the underlying UV completion. 
If their values lie close enough to the IR critical surface of the fixed point (spanned by its irrelevant directions), the RG flow is attracted towards the fixed point. Close to the fixed point, the RG flow becomes very slow and approximate scale symmetry is realized over a finite range of scales. Finally, the RG flow leaves the neighborhood of the fixed point along a relevant direction.

If there is no intermediate fixed-point regime, then a fundamental theory still provides initial conditions for the RG flow at some cutoff scale $\Lambda_{\rm{ QFT,\,} UV}'$, which lies below the Planck scale. Those initial conditions determine whether or not one enters a strong-coupling regime. This can be understood in terms of a single four-fermion coupling, which generically has the Gaussian and a non-Gaussian fixed point. The non-Gaussian fixed point constitutes the boundary of the basin of attraction of the Gaussian fixed point. An initial condition outside the basin of attraction necessarily enters a strong-coupling regime in the IR.

\begin{figure}
\begin{center}
\includegraphics[width=0.8\linewidth]{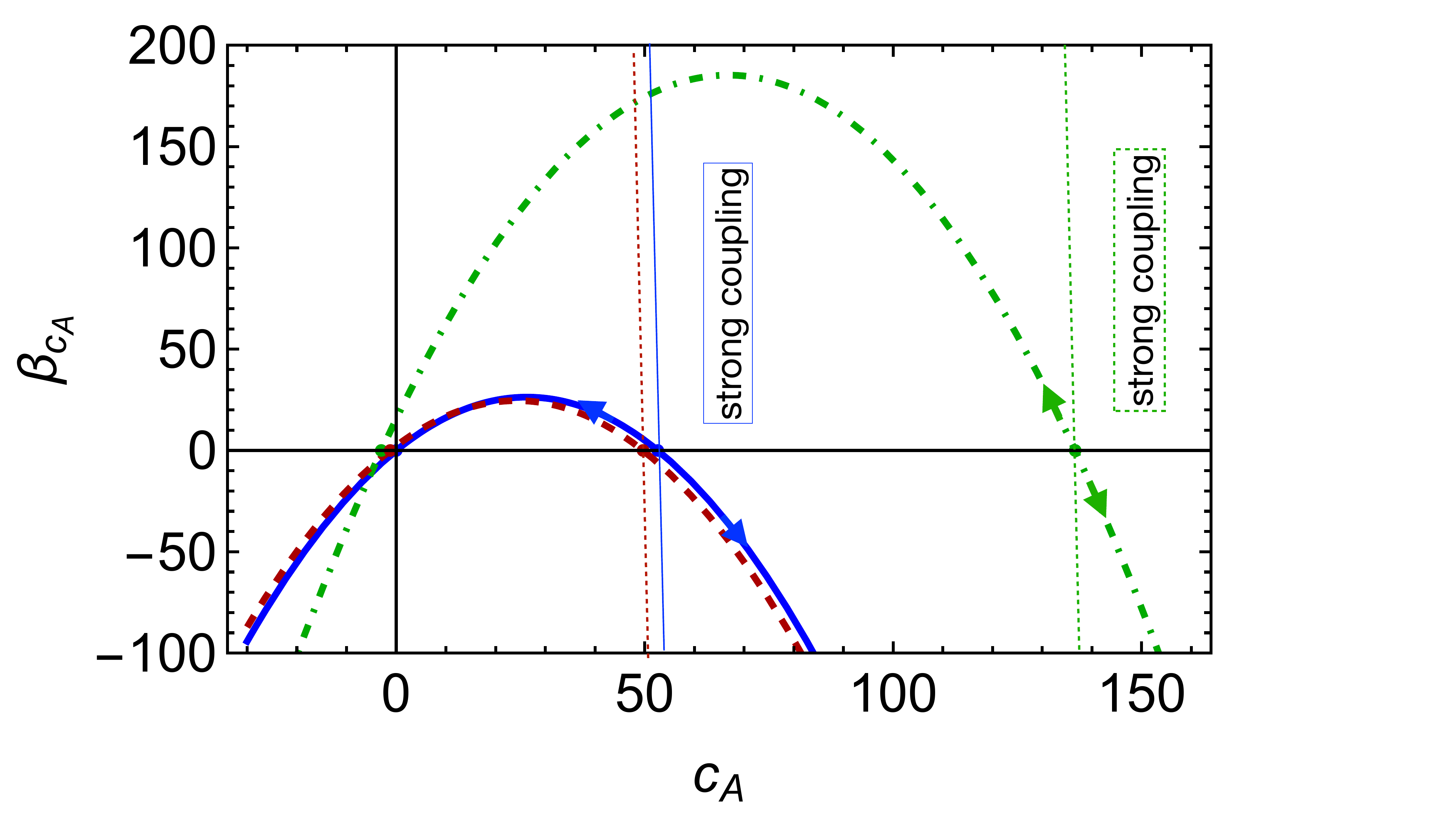}
\end{center}
\caption{\label{fig:betaca_illustration} We show the beta function for $c_A$ for $g=0$ (blue continuous line). All coupling values to the right of the interacting fixed point result in a strong-coupling regime in the IR. We also show two illustrative examples with non-zero $g$: the green dot-dashed beta function ($g=5,\, \lambda=0$ result in an increased regime that is safe from strong coupling. In contrast, the red dashed beta function ($g=500,\,\lambda=-20$) results in a very slight increase of the strong-coupling regime.}
\end{figure}

We now aim at understanding whether the volume of the basin of attraction of the (shifted) Gaussian fixed point increases or decreases under the effect of gravity. If it increases, we can say that gravity actually counteracts the formation of bound states (rather than just not triggering it). To this end, we determine the value of the second fixed point of $\beta_{c_A}$, which is relevant. Any coupling value larger than this fixed-point value results in an RG flow towards a strong-coupling regime in the IR, where we expect bound-state formation, see Fig.~\ref{fig:betaca_illustration}.

Our gravitational parameter space is in general divided into two regions: One region, in the vicinity of $\lambda =0$, features sizable gravitational effects. In the other region, for $\lambda$ sufficiently negative, gravitational effects are strongly suppressed. Therefore huge values of $g$ are needed in this region to achieve a sizeable impact of quantum gravity.

\begin{figure}
\begin{center}
\includegraphics[width=0.56\linewidth]{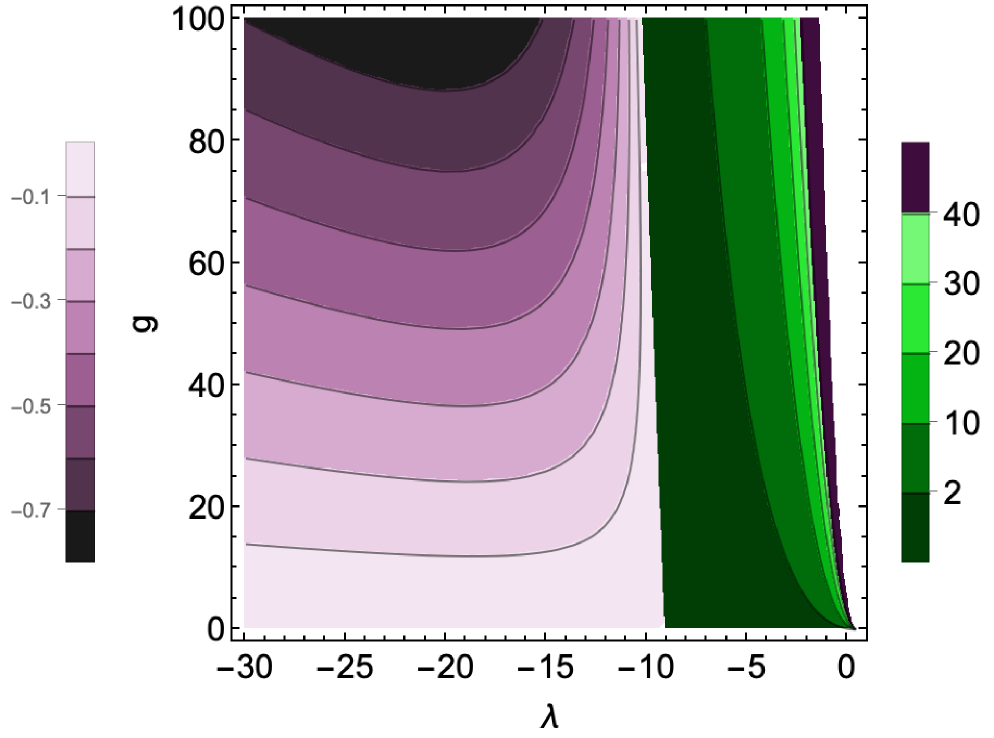}
\end{center}
\caption{\label{fig:glambdacA} We show the value of $c_{A\,\ast}- \frac{16\pi^2}{3}$, where the latter value is the fixed-point value in the absence of gravity. Where this quantity is positive, (green shaded region), a correspondingly larger interval of UV values of $c_A$ is safe from bound-state formation in the IR. Conversely, where this quantity is negative (fuchsia shaded region), a smaller interval is safe from the strong-coupling regime. We highlight that the values are much larger in the green-shaded than in the fuchsia-shaded region. Therefore, in practise, in the fuchsia-shaded region, the interval of values of $c_A$ that is safe from strong coupling is approximately the same as in the case without gravity.}
\end{figure}

We find that in the first region, gravitational fluctuations increase the range of values of couplings in which the system is safe from strong coupling and bound-state formation, cf.~Fig.~\ref{fig:glambdacA}. 
There is a critical curve, $\lambda_{\rm crit}(g)$ (near-horizontal in Fig.~\ref{fig:glambdacA}), beyond which gravitational fluctuations decrease the size of the safe range of coupling values.
This critical curve lies within the region in which the quantitative impact of gravitational fluctuations is small. 
Therefore, the absolute increase of the unsafe interval of values of $c_A$ is small. 
We conclude that gravitational fluctuations, where they have a sizable impact on the system, counteract bound-state formation and lead to a larger range of values of $c_A$ which do not flow into a strong-coupling region in the IR.

\section{Ordering principle for SMEFT coefficients in asymptotic safety}
\label{sec:categories}

Our analysis of a particular set of four-fermion interactions, together with previous work on higher-order interactions, most importantly \cite{Eichhorn:2017eht} and the specific examples in \cite{Christiansen:2017gtg,deBrito:2021pyi,Eichhorn:2021qet,deBrito:2021akp,Eichhorn:2023jyr, Assant:2025gto} encourage us to make general hypotheses about SMEFT interactions that we expect to hold even for those cases which have not been explicitly investigated.\\
The SMEFT is in general based on the assumption that the ordering of operators occurs on the basis of their mass-dimension. In asymptotic safety, the scaling exponents are no longer equal to the mass-dimension, because the scaling exponents at an interacting fixed point are shifted away from the mass-dimension due to quantum fluctuations,see, e.g., Eq.~\eqref{eq:thetaMSAS}. Thus, it is not a priori clear whether or not asymptotic safety results in an ordering principle for SMEFT coefficients that is very different from the usually assumed one. 
In turn, the ordering principle contains information on detectability of SMEFT coefficients. Thus, we discuss which of the following alternatives is most likely realized for the SMEFT as a whole:
\begin{itemize}
\item {\bf Alternative a}: Asymptotic safety changes the ordering principle for SMEFT coefficients significantly. The mass dimension of an interaction no longer provides a good guiding principle to determine the expected order of magnitude of a Wilson coefficient. In other words, asymptotic safety produces ``unnatural'' values for SMEFT coefficients.
\item {\bf Alternative b}: The mass dimension provides a largely robust guiding principle; only some individual SMEFT coefficients, singled out by an appropriate selection principle, may deviate from it. In this case, there are only individual SMEFT coefficients which can be ``unnatural''.
\item {\bf Alternative c:} The mass dimension works reliable as a guiding principle. Accordingly, the ordering of SMEFT coefficients in asymptotic safety obeys the EFT ordering. In other words, asymptotic safety provides a fully ``natural'' UV completion.
\end{itemize}

We point out that for purely gravitational interactions, there is good evidence that alternative c is realized \cite{Falls:2013bv, Falls:2014tra, Falls:2017lst, Falls:2018ylp,Kluth:2020bdv, Kluth:2022vnq}.
To arrive at a generalization of our results for four-fermion interactions, complemented by previous results on various higher-order couplings \cite{Eichhorn:2011pc, Meibohm:2016mkp, Eichhorn:2017eht, Christiansen:2017gtg, Eichhorn:2019yzm,deBrito:2021pyi, Laporte:2021kyp, deBrito:2021akp,Eichhorn:2021qet, Eichhorn:2023jyr,Eichhorn:2024wba, Assant:2025gto}, we first discuss the UV fixed-point regime and subsequently the flow to the IR.

\subsection{UV regime: Vanishing and non-vanishing SMEFT coefficients}
\label{sec:UV-regime}

We assume that the fixed-point properties are determined by the maximal symmetries of the kinetic terms, see Sec.~\ref{sec:globalsymmetries}. This hinges on two assumptions. First, we assume that the couplings of all marginal and relevant interactions in the SM can be set to zero at the fixed point.\footnote{There is also the possibility that gravitational contributions to the SM couplings have the right size to generate interacting fixed points for these \cite{Eichhorn:2018whv, Eichhorn:2025sux}. 
One would then need to consider the maximal symmetry given by the kinetic as well as such non-zero operators.
We leave it for the future to work out the detailed consequences for the SMEFT coefficients.} Second, we assume that gravitational fluctuations respect global symmetries. Hence, any interaction generated from gravity-matter vertices that are constructed from the kinetic terms must respect the symmetry of the kinetic terms.

Thus, any interaction which shares the maximal symmetry of the kinetic terms has a non-vanishing fixed-point value. All other interactions have vanishing fixed-point values.
To see this mechanism in action within the SMEFT, we identify the maximal symmetry.
For the case of vanishing SM interactions in the asymptotically safe regime, we can read it off by the symmetries of the kinetic terms of the SM.
These are
\begin{itemize}
    \item $\mathrm{O}(4)$ and shift symmetry for the Higgs field,
    \item $\mathrm{O}(12)$ and shift symmetry for the gauge fields, grouping all (decoupled) gauge fields into $A_\mu = (B_\mu,W^1_\mu,W^2_\mu,W^3_\mu,G^1_\mu,\dots,G^8_\mu)$. 
    \item $\mathrm{U}(N_W)$ for the fermionic field $\psi^I = (q_a, u^c, d^c,\dots)$ bundling all\footnote{$N_W$ counts all Weyl fermions, treating different color and weak isospin charges as separate elements of $\psi^I$.
        $N_W$ evaluates to $15$ per generation in the SM and $16$ per generation if right-handed neutrinos are included.} left-handed and charge conjugated right-handed fermions.
\end{itemize}
Shift symmetry is a remnant of gauge symmetry at vanishing gauge coupling. We stress that for the $\mathrm{SU}(N)$ non-Abelian gauge fields, the same symmetry holds, because, at vanishing gauge coupling, only an abelian part of the group remains and the global symmetry of the kinetic term becomes that of $N^2-1$ independent vector fields for each gauge field.\footnote{We stress that this is an unusual limit for a gauge theory. It is the limit in which the gauge fields ``forget'' that they are associated to the adjoint representation of a Lie algebra. Therefore, the remaining part of their kinetic term is an \emph{Abelian} version of a field strength, $F_{\mu\nu} = \partial_{\mu}A_{\nu}- \partial_{\nu}A_{\mu}$.}

Reduced to maximal symmetry of the MSAS regime, the kinetic terms read
\begin{equation}
    \label{eq:L_MSAS_4}
    \mathcal{L}_{\text{MSAS}}^{D=4} = -\frac{1}{4} F_{\mu\nu}^I F_I^{\mu\nu} 
    + \psi^\dagger_J \barsigma^\mu \nabla_\mu \psi^J
    + \partial_{\mu} \varphi_a^\dagger \partial^{\mu} \varphi^a,
\end{equation}
with $F_{\mu\nu}^I = \nabla_\mu A_\nu^I - \nabla_\nu A^I_\mu$ (which holds also for Yang--Mills theory at vanishing coupling). 
Without the gravitational coupling, the Lagrangian in Eq.~\eqref{eq:L_MSAS_4} describes the completely free part of the Standard Model.
These terms constitute the only building blocks of the MSAS theory subspace.
See Sec.~\ref{sec:msas_toymodel} for the construction in our toy model.
The kinetic terms in Eq.~\eqref{eq:L_MSAS_4} are not completely free, once gravity is considered, because gravity does not decouple. Instead, the action built from Eq.~\eqref{eq:L_MSAS_4} contains a factor $\sqrt{-g}$ in the integration measure over spacetime, the indices in Eq.~\eqref{eq:L_MSAS_4} are contracted with the metric and the derivative $\nabla_{\mu}$ acquires a spin connection when acting on $\psi^I$ (see App.~\ref{app:weyl_fermions_in_qg}).
In contrast, the kinetic terms for scalars and vectors do not contain the connection, because the covariant derivative does not require a connection when acting on spacetime scalars, and the connection, while present in $\nabla_{\mu}A_{\nu}^I$, cancels in $F_{\mu\nu}^{I}$. 
From these occurrences of the metric and the connection (which itself depend on the metric), vertices that couple the matter field to gravitational fluctuations arise. In turn, these provide us with diagrams with any number of external matter legs. The resulting beta functions generate non-zero couplings for those SMEFT interactions which share the above global symmetries.

We can now start to categorize higher order operators. We distinguish those that are zero from those that are non-zero at the fixed point in the MSAS. To this end, we work our way through the lowest-order mass dimension-interactions in the SMEFT.

At mass-dimension 5,  the SMEFT contains only the Weinberg operator, 
\begin{equation}
    \Gamma_k^{\text{Weinberg}} = \levicivita_{ab} \levicivita_{cd} \levicivita^{\alpha\beta}(l_\alpha^a \varphi^b) (l_\beta^c \varphi^d) + \hc
\end{equation}
for which there are indications that it is irrelevant and zero at the Planck scale in asymptotic safety \cite{deBrito:2025ges}.\footnote{
While neutrino masses can likely not arise in asymptotic safety through the Weinberg operator at a fundamental level, there is actually a dynamical mechanism through which neutrino Yukawa couplings can remain small \cite{Held:2019vmi, Kowalska:2022ypk, Eichhorn:2022vgp, Eichhorn:2025sux}. This provides a mechanism for neutrino Dirac mass generation that results in realistic neutrino masses without the need for extra heavy degrees of freedom. More standard, see-saw type mechanisms can, however, also likely be accommodated in asymptotic safety \cite{Domenech:2020yjf}, based on the fact that a Majorana mass term remains a relevant coupling \cite{DeBrito:2019rrh}, although with an upper bound on the seesaw scale \cite{deBrito:2025ges}.
The required extension of the SM would of course introduce additional SMEFT operators not considered in our analysis.}
It is also not maximally symmetric, as the term breaks the shift symmetry of the Higgs field.

At mass-dimension 6 there are 3045 real operators \cite{Alonso:2013hga, Alonso:2014zka}\footnote{Neglecting baryon number violating terms would bring this count to 2499.}, and the number of operators `explodes' for higher mass dimensions \cite{Henning:2015alf}.
Are these all present in the asymptotically safe Standard Model?

Adopting the convention of \cite{Grzadkowski:2010es}, the SMEFT mass-dimension-6 operators can be divided into 8 classes. Using $X$ as a general label for gauge fields, $\varphi$ for the Higgs field, $\psi$ for fermions and $D$ for covariant derivatives, these may be abbreviated as $X^3$, $\varphi^6$, $\varphi^4 D^2$, $\psi^2 \varphi^3$, $X^2\varphi^2$, $\psi^2 X \varphi$, $\psi^2\varphi^2 D$ and $\psi^4$ (Tab.~\ref{tab:smeft_dim_6}).
The class of four-fermion operators $\psi^4$ can further be subdivided by chirality into the following 5 subclasses, $(L^\dagger\! L)(L^\dagger\! L)$, $(R^\dagger\! R)(R^\dagger\! R)$, $(L^\dagger\! L)(R^\dagger\! R)$, $(L^\dagger\! R)(L^\dagger\! R)$ and $B$-violating operators.\footnote{The gauge and Higgs sector of the SMEFT constitute $4+3+6+8+16+9=46$ or $4+3+8+54+144+81 = 294$ operators for $1$ or $3$ generations respectively. The four-fermion sector count $5+7+8+10+8=38$ or $297 + 450 + 648 + 810 + 546 = 2751$ operators for 1 or 3 generations of matter \cite{Grzadkowski:2010es, Alonso:2013hga, Alonso:2014zka}.}

Of these classes of operators, the class $X^3$ violates the $\mathrm{O}(12)$ symmetry of the kinetic terms of the gauge fields and can thus not be present at the SGFP.
Similarly, the operator classes $\varphi^6$, $\varphi^4 D^2$, $\psi^2 \varphi^3$, $X^2 \varphi^2$, $\psi^2 X \varphi$ and $\psi^2\varphi^2 D$ violate shift-symmetry of the Higgs field; some of them violate additional symmetries. We conclude that none of these operators can be present at the SGFP. Therefore, no interaction in the gauge and Higgs sectors of Tab.~\ref{tab:smeft_dim_6} is underlined, meaning that none of the corresponding couplings has a non-vanishing fixed-point value.

In the four-fermion operator category, operators of mixed chirality, i.e.\ $(L^\dagger\! R)(R^\dagger\! L)$ and $(L^\dagger\! R)(L^\dagger\! R)$, or baryon-number violating type cannot be induced by gravity coupling to the kinetic term.\footnote{One can even show that baryon-number violating operators which would mediate proton decay become more irrelevant under the impact of quantum gravity \cite{Eichhorn:2023jyr}, resulting in the expectation that experiments looking for proton decay should only push the lower bound on the proton lifetime to larger values, but not detect any decays.}
However, some of the four-fermion operators of paired chirality, i.e.\ $(L^\dagger\! L)(L^\dagger\! L)$, $(R^\dagger\! R)(R^\dagger\! R)$, $(L^\dagger\! L)(R^\dagger\! R)$ can be induced depending on their flavor structure.

To satisfy maximal symmetry, the bilinears must be singlets of all three gauge groups. 
This restriction excludes all octet and triplet current operators.
Moreover, the kinetic terms conserve flavor, which is implicit in $\mathrm{U}(N_W)$ and therefore, all flavor mixing operators must be excluded.
For instance, the $(L^\dagger\! L)(R^\dagger\! R)$ operator $[O_{le}]^{prst} = (l_p^\dagger \barsigma^\mu l_r)(e^\dagger_s \sigma^\mu e_t)$ is induced only for $p=r$ and $s=t$.\footnote{
For $(L^\dagger\! L)(L^\dagger\! L)$ and $(R^\dagger\! R)(R^\dagger\! R)$, due to the Fierz identities $\big(\chi^\dagger_1 \sigma^\mu \chi_2\big) \big(\chi^\dagger_3 \sigma^\mu \chi_4\big) = \big(\chi^\dagger_1 \sigma^\mu \chi_4\big) \big(\chi^\dagger_3 \sigma^\mu \chi_2\big)$ and $	\big(\psi^\dagger_1 \barsigma^\mu \psi_2\big) \big(\psi^\dagger_3 \barsigma^\mu \psi_4\big) = \big(\psi^\dagger_1 \barsigma^\mu \psi_4\big) \big(\psi^\dagger_3 \barsigma^\mu \psi_2\big)$, not all combinations of flavor indices $q$, $r$, $s$, $t$ are independent. 
Whether an operator actually mixes flavors may thus not be immediately visible.}

We conclude that the following operators of the mass-dimension-6 SMEFT are asymptotically safe for the maximally symmetric shifted Gaussian fixed point scenario given that $p=r$ and $s=t$,
\begin{align}
    (L^\dagger\! L)(L^\dagger\! L):&\quad
    O_{ll}\,,
    O_{qq}^{(1)}\,,
    O_{lq}^{(1)}\,, \nonumber\\
    (R^\dagger\! R)(R^\dagger R):&\quad
    O_{ee}\,,
    O_{uu}\,,
    O_{dd}\,,
    O_{eu}\,,
    O_{ed}\,,
    O_{ud}^{(1)}\,, \nonumber\\
    (L^\dagger\! L)(R^\dagger R):&\quad
    O_{le}\,,
    O_{lu}\,,
    O_{ld}\,,
    O_{qe}\,,
    O_{qu}^{(1)}\,,
    O_{qd}^{(1)}\,.
    \label{eq:list_msas_operators}
\end{align}
All others vanish at the fixed point. 
The interactions listed in Eq.~\eqref{eq:list_msas_operators} are easily constructible from the minimal kinetic Lagrangian in Eq.~\eqref{eq:L_MSAS_4} and have the structure of Eq.~\eqref{eq:4F_action_msas}.
We expect this categorization and identification of non-zero operators to be straightforwardly generalizable to higher-mass-dimension operators.

In summary, based on this symmetry consideration, only the underlined terms in Tab.~\ref{tab:smeft_dim_6} are non-vanishing at the UV fixed points, all other SMEFT couplings vanish at the fixed point, at least if all marginal SM interactions have vanishing fixed-point values.

To close the discussion of the UV fixed-point properties, let us briefly highlight the potential consequences of a more complex fixed-point structure.
The interacting UV fixed point might feature SM couplings (canonically marginal) that are not asymptotically free at the fixed point.
In particular, the Abelian gauge coupling \cite{Harst:2011zx, Eichhorn:2017lry} or the top or bottom Yukawa coupling may be non-zero \cite{Eichhorn:2017ylw, Eichhorn:2025sux}.
In this case, the new non-zero operator helps span a larger hypersurface in theory space that reduces the maximal symmetry.
This induces further non-zero SMEFT couplings through purely gravitational loops.\footnote{In a standard discussion of asymptotic safety, the full effective action is constructed, in which SM and gravitational degrees of freedom are all integrated out. However, the SMEFT is constructed under the assumption that only BSM degrees of freedom are integrated out and SM fluctuations, no matter how high their momenta, are not. Thus, a comparison to the SMEFT requires us to consider only a subset of those diagrams that would be included in the calculations of the full effective action from asymptotic safety.}
Operators of the maximally symmetric subspace are still induced solely by the interactions of the kinetic terms with gravity and thus only dependent on the gravitational parameters.
However, in the extended setting, the additional operators are excited due to the inclusion of a non-vanishing SM interactions in the UV.
This means that the induction term in the corresponding beta function must be proportional to that SM coupling, which now parametrically controls the induced operator's strength in the UV. 
For a study in a toy model, accounting for gravity and a dynamical, quantized gauge field, see \cite{deBrito:2023kow}.
For the present considerations, we will restrict ourselves to the maximal symmetry given by the kinetic terms.

\subsection{IR regime: Vanishing and non-vanishing SMEFT coefficients}
\label{sec:IR-regime}

The IR values of couplings are largely independent of their fixed-point values only if a given coupling is relevant. Based on the present work and the previous results in various sectors \cite{Narain:2009fy,Eichhorn:2011pc, Meibohm:2016mkp, Eichhorn:2017eht, Christiansen:2017gtg, Eichhorn:2017als,Eichhorn:2019yzm,deBrito:2021pyi, Laporte:2021kyp, deBrito:2021akp,Eichhorn:2021qet, Eichhorn:2023jyr,Eichhorn:2024wba, Assant:2025gto}, we conjecture that all induced interactions remain irrelevant.
This is based on two observations: First, for all fields except for fermions, the induced interactions are at least order 8 in mass dimension.
A shift into relevance would be in tension with the near-perturbative nature of the asymptotically safe fixed point, which we regard as rather well-established \cite{Falls:2013bv, Falls:2014tra, Falls:2017lst, Falls:2018ylp,Eichhorn:2018akn,Eichhorn:2018nda,Eichhorn:2018ydy, Kluth:2020bdv, Kluth:2022vnq}, see also \cite{Eichhorn:2026uqj} for a discussion of the evidence.
Second, induced four-fermion interactions have been analyzed in multiple settings \cite{Eichhorn:2011pc, Eichhorn:2017eht, deBrito:2020dta, deBrito:2023kow} including this work, featuring representative substitutes for most classes of four-fermion SMEFT operators, with no concrete indications for relevance in near-perturbative settings.

Thus, based on our discussion for four-fermion couplings, we can conjecture that couplings outside the maximally symmetric subspace for the SMEFT are the strongest candidates for couplings that can circumvent Planck-scale suppression. These are all interactions that are \emph{not} underlined in Tab.~\ref{tab:smeft_dim_6}. 
We caution that at least in our study, the required fixed-point collision that renders these couplings relevant only occurs at very large values of $g_{\ast}$, which are not close to fixed-point values typically found in calculations.
Whether or not such fixed-point collisions are robust under extensions of the truncation needs to be studied along similar lines as in \cite{deBrito:2023myf}.
We note that the fact that the shifted Gaussian fixed point for the four-fermion interactions remains largely perturbative even after the fixed-point collision, could be interpreted as a sign of robustness -- at least for the truncation of the matter sector -- of our results.

Coming back to the three alternatives discussed in the beginning of this section, we conclude that {\bf alternative a} is implausible, given our results.
{\bf Alternative b} cannot be excluded, but, at least in our setting, requires very large fixed-point value of $g_{\ast}$. Thus, {\bf alternative c} remains as the most plausible case, based on our analysis. 
While potentially somewhat disappointing from an observational point of view, this result is well in line with previous results that support the near-perturbative nature of the asymptotically safe fixed point.
In particular, we conclude that the ordering of interactions into relevant and irrelevant at the SGFP follows the EFT ordering principle; i.e., canonical power counting is a robust guiding principle.

From a technical point of view, our result supports the robustness of truncations of the Wetterich equation that include operators based on an ordering by mass dimension. This is crucial for the reliability of existing results for the marginal SM couplings, summarized in \cite{Eichhorn:2026uqj}.

\section{Conclusions and outlook}
\label{sec:conclusions}

The SMEFT constitutes a natural framework to investigate the interplay of asymptotically safe quantum gravity with matter. This is because the ``landscape'' of asymptotic safety with SM matter can be charted by the SMEFT coefficients, in addition to non-minimal matter-curvature couplings.\footnote{This pertains to the coefficients of local terms. In addition, the full effective action that arises from integrating out metric fluctuations is expected to contain non-local terms. These are, however, unlikely to be dynamically relevant at typical particle-physics scales.}

We can thus address both structural as well as phenomenological questions about asymptotic safety within the framework of the SMEFT.

At the structural level, the SMEFT coefficients provide the full set of possible interactions that may in principle be generated at an asymptotically safe fixed point.
In the SMEFT, these are provided order by order in canonical power counting.
At a near-perturbative fixed point, canonical power counting provides a good guiding principle for the distinction into relevant and irrelevant perturbations at the fixed point.
Therefore, we expect to be able to proceed order by order in the SMEFT coefficients to determine whether an asymptotically safe fixed point of gravity with SM matter exists and what its properties are.
In this vein, parts of the SMEFT have been studied, e.g., dimension-eight-interactions field-strength-terms, which are the leading-order interactions in the gauge sector \cite{Christiansen:2017gtg,Eichhorn:2019yzm,Eichhorn:2021qet, Eichhorn:2024wba, Knorr:2024yiu}, as well as dimension-eight scalar derivative interactions \cite{Eichhorn:2012va, deBrito:2021pyi, Laporte:2021kyp, deBrito:2023myf}, which are the leading-order terms in the Higgs sector.
For the fermion sector, an important aspect was neglected in all previous work on dimension-six four-fermion interactions \cite{Eichhorn:2011pc, Meibohm:2016mkp, Eichhorn:2017eht, deBrito:2020dta, deBrito:2023kow}: the chiral nature of the SM fermion sector, in which the left-handed fermions satisfy different gauge symmetries (here simplified to global symmetries) from the right-handed fermions. 
Here, we take a crucial step towards a realistic treatment of the SM fermion sector, by investigating the four-fermion interactions of a left-handed quark doublet combined with two right-handed singlets. 
We limit ourselves to a single generation of electro-weakly charged quarks. 
This reduces the several thousands of four-fermion interactions in the SMEFT to seven.

Our main results are that within this space of seven four-fermion couplings, a suitable fixed point exists for a UV completion under the impact of quantum gravity. At this fixed point, six of the four-fermion couplings are non-zero.
We find that these can be mapped to the fixed-point values found in \cite{Eichhorn:2011pc}, which was based on Dirac fermions.
In fact, our results and the results in \cite{Eichhorn:2011pc} can be rephrased in terms of a highly symmetric setting, with a $U(N_W)$ symmetry that maps all Weyl fermions into each other.
Therefore, the non-zero fixed-point values of the couplings $c_{qq}, c_{qu}, c_{qd}, c_{uu}, c_{dd}, c_{ud}$, all follow from a fixed point of a single $U(N_W)$ invariant four-fermion interaction.\footnote{On this basis, we also explain why \cite{Eichhorn:2011pc} found that gravitational fluctuations acting on a system with $N_F$ Dirac fermions induce the four-fermion interaction $(\bar{\Psi}^i\gamma_{\mu}\gamma_5\Psi^i)^2$, but not $(\bar{\Psi}^i\gamma_{\mu}\Psi^i)^2$, even though the two constitute a basis of  $SU(N_F)_L\times SU(N_F)_R$-symmetric four-fermion interactions.
However, from the point of view of gravity, the fermions can actually all be organized into a $2N_F$-component Weyl fermion.
The induced four-fermion interaction therefore is based on the axial current only.}
The existence of this fixed point constitutes another non-trivial check of the asymptotically safe Standard Model.

At the phenomenological level, we ask whether asymptotic safety necessarily follows the EFT ordering principle. Specifically, we take this ordering principle to mean that all SMEFT interactions are suppressed by the same UV scale and the dimensionless Wilson coefficients are roughly $\mathcal{O}(1)$, i.e., not so large or small that the ordering scheme arising from the dimensionality of the interactions is disturbed. 
We find that there is a mechanism which can in principle circumvent this ordering principle.
It consists in a fixed-point collision that involves $c_{quqd}$ in a crucial role.
This coupling is the only one that is strictly zero at the most predictive fixed point which exhibits the (non-manifest) $U(N_W)$ symmetry.
The eigendirection associated to it is irrelevant at small $g_{\ast}$, but can become relevant at large $g_{\ast}$.
This is a consequence of a fixed-point collision, reminiscent of stability trading of fixed points in statistical physics, and similarly shaped by the underlying symmetries of distinct fixed points.

Abstracting from the results in the subsector of the SMEFT that we study, we conjecture that: SMEFT couplings which respect the maximal global symmetries of the kinetic terms of the matter fields (with gauge and Yukawa interactions as well as scalar potential set to zero) are non-zero at the fixed point.
These are underlined in Tab.~\ref{tab:smeft_dim_6}.
They remain irrelevant and thus follow EFT ordering in the IR, with dimensionless SMEFT coefficients of $\mathcal{O}(1)$ and the suppression scale set by $M_{\rm Planck}$ with the canonical dimension of the term.
In contrast, the other couplings vanish at the maximally symmetric fixed points and are not underlined in Tab.~\ref{tab:smeft_dim_6}.
These non-underlined terms are candidates for \emph{relevant} interactions, if they exhibit the same stability-trading mechanism that $c_{quqd}$ does in our example.
Therefore, from a phenomenological perspective, these interactions are the most interesting ones, because they may potentially circumvent Planck-scale suppression.
We caution that for our system, the stability-trading mechanism requires large values of $g_{\ast}$ that are not usually realized by the gravitational fixed point. 

Overall, we therefore strengthen the indications that an asymptotically safe fixed point for gravity with the Standard Model not only exists, but also follows a relatively strict canonical ordering principle.
This evades ``exotic'' phenomenological scenarios, in which higher-order interactions feature huge coefficients. 
This is of course well in line with the existing bounds on SMEFT coefficients from the LHC \cite{ATLAS:2024lyh,CMS:2025ugn}.
It additionally implies that even in dedicated settings which probe much higher suppression scales than is generally reachable at the LHC (as is possible for specific dimension-five interactions in the SMEFT and beyond, such as those resulting in proton-decay), the asymptotically safe prediction can be expected to be that couplings are Planck-scale suppressed.

This result also provides a robust principle to organize calculations in.
Without having to go to very high orders in the SMEFT expansion -- which is technically unfeasible due to the fast growth of the number of interactions -- we can expect that no dimension-eight interaction becomes relevant and very likely not even dimension-six interactions can become relevant.
 
We note that the fixed-point structure for the four-fermion interactions results in the prediction of precise relations between the irrelevant four-fermion couplings in the IR, namely Eq.~\eqref{eq:MSAS_relation_for_c}.
These relations hold by virtue of the fixed-point values.
Because the critical exponents are always negative (and can become large in absolute value for large enough $g_{\ast}$), the Planck-scale value of the couplings always corresponds to the fixed-point values, even if the RG flow does not start at the fixed point, but only somewhere within its basin of attraction.
This is an example of the predictive power of asymptotic safety: it gives rise to very specific relations between couplings associated to irrelevant interactions. 
We expect that based on the principle that the maximal global symmetry of the kinetic terms determines the fixed-point values, we expect that analogous relations determine, e.g., the various interactions involving different gauge fields.
As discussed above these predictions are subject to Planck-scale suppression and thus do not appear testable in practise. 
Nevertheless, these relations may constitute important \emph{theoretical} tests of asymptotic safety: similar predicted relations between higher-order coefficients for vectors are subject to IR causality constraints that constrain the dimensionless ratios of SMEFT coefficients, irrespective of the overall suppression scale.
Such relations can be tested in asymptotic safety, see \cite{Eichhorn:2024wba, Knorr:2024yiu}. 
For fermions, analogous relations from considerations of positivity are more tricky to arrive at; and constraints on dimension-six-interactions in general require additional assumptions, see, e.g., \cite{Remmen:2020uze}.
Nevertheless, sum rules for dimension-six four-fermion interactions may be compared to asymptotically safe results in the future.

As an outlook, we add that, despite being irrelevant and thus not of phenomenological importance, non-vanishing four-fermion couplings can constitute an important quantitative correction to fixed-point properties of canonically marginal interactions. 
For instance, intriguing phenomenology has been extracted by considering flavor-universal gravitational contributions to Yukawa couplings \cite{Eichhorn:2017ylw, Eichhorn:2018whv, Eichhorn:2025sux}. We expect that the flavor-universality can be broken at a subleading level due to four-fermion interactions. 
Specifically, the scenario we have in mind is the following: In the settings in \cite{Eichhorn:2017ylw, Eichhorn:2018whv, Eichhorn:2025sux}, an asymptotically safe fixed point can feature a nonzero top-Yukawa coupling. 
In turn, such a coupling contributes to four-fermion couplings which involve the top quark. These take fixed-point values different from the four-fermion couplings which only include the other quarks. 
Finally, four-fermion couplings contribute to the beta-function of the Yukawa couplings and constitute an indirect mechanism to break the flavor-universality of gravitational effects. 
We expect, in line with the results in \cite{Eichhorn:2017eht} that these effects are subleading compared to the direct gravitational contribution \cite{Oda:2015sma,Eichhorn:2016esv, Eichhorn:2017eht}. 
Therefore, these effects might constitute a quantitative correction to results of phenomenological studies such as \cite{Eichhorn:2017ylw, Eichhorn:2018whv, Eichhorn:2025sux}, and complement recent efforts to reduce and quantify systematic uncertainties in gravity-matter systems \cite{Riabokon:2025ozw, deBrito:2025nog}.

\subsubsection*{Acknowledgements}

We thank Marc Schiffer and Benjamin Knorr for discussions. M.~G.~is grateful to the University of the Faroe Islands for hospitality.
A.~E.~and M.~G.~acknowledge the European Research Council's (ERC) support under the European Union’s Horizon 2020 research and innovation program Grant agreement No.~101170215 (ProbeQG). 

\bibliographystyle{JHEP}
\bibliography{refs}

\appendix

\section{Weyl fermions in quantum gravity}
\label{app:weyl_fermions_in_qg}

Some parts of our study employ tools developed for  fermions on curved spacetime. Most importantly, this includes the treatment of the tangent space: we employ the spin-base invariant formalism introduced by Weldon \cite{Weldon:2000fr} which was generalized by Lippoldt \cite{Gies:2013noa, Lippoldt:2015cea}. The results are fully equivalent to using an O(4) gauge fixing \cite{Woodard:1984sj, vanNieuwenhuizen:1981uf} that relates vielbein fluctuations to metric fluctuations, as in \cite{Eichhorn:2011pc}.
 We refer to \cite{Dreiner:2008tw} for an introduction to two-component notation introduced by Infeld and van der Waerden.
Weyl fermions on curved spacetimes are discussed extensively in \cite{Penrose:1986ca}.
We summarize here the main points.

The generators of the $(\frac{1}{2},0)$ and $(0,\frac{1}{2})$ representations of the Lorentz group,  in which Weyl fermions  transform, are 
\begin{align}
	(\sigma^{ab})_\alpha{}^\beta &= \frac{\ii}{4}\left(\sigma^a_{\alpha\dotgamma} \barsigma^{b\,\dotgamma\beta} - \sigma^b_{\alpha\dotgamma} \barsigma^{\mu\,\dotgamma\beta}\right)\,, \\ 
	(\barsigma^{ab})^\dotalpha{}_\dotbeta &= \frac{\ii}{4}\left(\barsigma^{a\,\dotalpha\gamma} \sigma^{b}_{\gamma\dotbeta} - \barsigma^{b\,\gamma\dotbeta} \sigma^{b}_{\gamma\dotbeta}\right) \,,
\end{align}
where Greek indices starting with $\alpha, \beta,\dots$ are spinor indices, undotted for left-handed and dotted for right-handed spinors.
Latin indices denote flat Minkowski space.
In Lorentzian mostly minus signature, $\sigma^{\mu}= (\mathds{1}, \sigma_i)$, and $\bar{\sigma}^{\mu}=(\mathds{1}, -\sigma_i)$ with $\sigma_i$ ($i \in \{1,2,3\}$) denoting the Pauli matrices. 
In Euclidean signature $\sigma^\mu = (\mathds{1},-\ii \sigma_i)$ and $\barsigma^\mu = (\mathds{1}, \ii \sigma_i)$.
Lorentz transformations read 
\begin{equation}
    \begin{aligned}
	M &= M_{(\frac{1}{2},0)}{}_\alpha{}^\beta(\theta_{ab}) = \exp\left(\frac{\ii}{2} \theta_{ab} \sigma^{ab}\right)\,,
	 \\
	M^\dagger &= M_{(0,\frac{1}{2})} {}^\dotbeta {}_\dotalpha(\theta_{ab}) = \exp\left(\frac{\ii}{2} \theta_{ab} \barsigma^{ab}\right)\,,
    \end{aligned}
\end{equation}
noting $\left( (\sigma^{ab})_\alpha{}^\beta\right)^\dagger = (\barsigma^{ab})^\dotalpha{}_\dotbeta$ 
\begin{align}
    \psi_\alpha &\to M_\alpha{}^\beta \psi_\beta\,, &
    \psi^\alpha &\to \psi^\beta (M^{-1})_\beta{}^\alpha \,,\\
    \chi_\dotalpha &\to \chi_\dotbeta (M^\dagger)^\dotbeta{}_\dotalpha \,, &
    \chi^\dotalpha &\to ((M^{-1})^\dagger)^\dotalpha{}_\dotbeta \chi^\dotbeta\,,
\end{align}
where $\psi$ is a left-handed Weyl fermion and $\chi$ is a right-handed Weyl fermion.

Tetrads (vielbeins) connect a locally flat tangent space to general coordinates (with Greek indices starting with $\mu$)
\begin{equation}
    g_{\mu\nu}(x) = e^a{}_\mu(x) e^b{}_\nu(x) \eta_{ab}\,.
\end{equation}
Due to metric compatibility they satisfy
\begin{equation}
    \nabla_\mu g_{\rho\sigma} = 0 \implies \nabla_\mu e^a{}_\nu = \nabla_\mu e^a{}_\nu = \partial_\mu e^a{}_\nu - \Gamma^\lambda_{\nu\mu} e^a{}_\lambda + \omega_\mu{}^a{}_b e^b{}_\nu = 0\,,
\end{equation}
where $\omega_\mu{}^a{}_b$ is a connection for the local coordinates
\begin{equation}
    \omega_\mu{}^a{}_b = -e_b{}^\nu \left(\partial_\mu e^a{}_\nu - \Gamma^\lambda_{\mu\nu} e^a{}_\lambda\right)\,.
\end{equation}

The spin connection $B_\mu$ for an unspecified fermion $\Psi$
\begin{equation}
	\nabla_\mu \Psi(x) = \partial_\mu \Psi(x) + B_\mu(x) \Psi\,,
\end{equation}
 can be formulated in terms of $\omega$.
We can take the connection of the locally flat coordinates and evaluate it for the required representations
\begin{equation}
    B_\mu(x) = \ii\omega_\mu{}^{ab}(x) \Sigma_{ab}\,.
\end{equation}
As general results \cite{Parker:2009uva},
\begin{equation}
    \nabla_\mu \Psi = \partial_\mu \Psi + \ii \omega_\mu{}^{ab}\Sigma_{ab}\Psi\,,
\end{equation}
and
\begin{equation}
    [\nabla_\mu, \nabla_\nu] \Psi = -\ii R_{\mu\nu}{}^{ab} \Sigma_{ab} \Psi\,.
\end{equation}
For Dirac fermions, $\Sigma_{ab} = -\frac{\ii}{8} [\gamma_a,\gamma_b]$, while for left-handed Weyl fermions $\Sigma_{ab} = -\frac{1}{2}\sigma_{ab}$ and for right-handed Weyl fermions $\Sigma_{ab} = -\frac{1}{2}\barsigma_{ab}$.
We can upgrade the flat sigma matrices to include the local transformation 
\begin{equation}
    \sigma^\mu(x) = \sigma^a e_a{}^{\mu}(x)\,,\quad
    \barsigma^\mu(x) = \barsigma^a e_a{}^{\mu}(x)\,.
\end{equation}
Using 
\begin{equation}
	P_L \gamma^\mu = \gamma^\mu P_R \doteq \begin{pmatrix} 0 & \sigma^\mu \\ 0 & 0\end{pmatrix}, \quad
	P_R \gamma^\mu = \gamma^\mu P_L \doteq \begin{pmatrix} 0 & 0 \\ \barsigma^\mu & 0 \end{pmatrix},
  \label{eq:projectors}
\end{equation}
this can be put into Dirac form
\begin{align}
    \Sigma_{ab}^L &= -\frac{1}{2} \sigma_{ab} = -\frac{\ii}{8}\left(\sigma_a \barsigma_b - \sigma_b \barsigma_a\right) \stackrel{\wedge}{=} -\frac{\ii}{8} [\gamma_a,\gamma_b] P_L\,, \nonumber\\
    \Sigma_{ab}^R &= -\frac{1}{2} \barsigma_{ab} = -\frac{\ii}{8}\left(\barsigma_a \sigma_b - \barsigma_b \sigma_a\right) \stackrel{\wedge}{=} -\frac{\ii}{8} [\gamma_a,\gamma_b] P_R\,.
\end{align}

$\tilde{\levicivita}^{\mu\nu\rho\sigma}$ is the flat four-dimensional Levi-Civita symbol defined as 
\begin{equation}
	\label{eq:flat_levicivita}
	\tilde{\levicivita}^{\mu\nu\rho\sigma} = \begin{cases}
		+1 & \text{for even permutations of } \left\{\mu,\nu,\rho,\sigma\right\} \\
		-1 & \text{for odd permutations of } \left\{\mu,\nu,\rho,\sigma\right\}  \\
		0  & \text{otherwise}\,.
	\end{cases}
\end{equation}
For the Levi-Civita tensor for curved spacetime $\levicivita_{\mu\nu\rho\sigma}$ we need to include the density factor $\sqrt{g}$, such that we have
\begin{equation}
	\levicivita_{\mu\nu\rho\sigma} = \sqrt{g}\, \tilde{\levicivita}_{\mu\nu\rho\sigma}\,.
\end{equation}
This allows us to show that $\gamma^5$ (implicit in the projectors $P_L$ and $P_R$) is completely independent of spacetime curvature.
\begin{equation}
	\label{eq:gamma5_spacetime_independence}
	\begin{aligned}
		\gamma^5 =&\; - \frac{1}{4!} \levicivita_{\mu\nu\rho\sigma} \gamma^\mu \gamma^\nu \gamma^\rho \gamma^\sigma \\
		=&\; - \frac{1}{4!} \sqrt{\det(g)}\, \tilde{\levicivita}_{\mu\nu\rho\sigma} e_a^\mu e_b^\nu e_c^\rho e_d^\sigma\, \gamma^a \gamma^b \gamma^c \gamma^d \\
		=&\; - \frac{1}{4!} \sqrt{\det(\delta_{ab} e_\mu^a e_\nu^b)}\, \tilde{\levicivita}_{\mu\nu\rho\sigma} e_1^\mu e_2^\nu e_3^\rho e_4^\sigma \tilde{\levicivita}_{abcd}\, \gamma^a \gamma^b \gamma^c \gamma^d \\
		=&\; - \frac{1}{4!} \det(e_\mu^a) \det(e_a^\mu)\, \tilde{\levicivita}_{abcd}\, \gamma^a \gamma^b \gamma^c \gamma^d \\
		=&\; - \frac{1}{4!} \levicivita_{abcd}\, \gamma^a \gamma^b \gamma^c \gamma^d.
	\end{aligned}
\end{equation}
One can thus use Eq.~\eqref{eq:projectors} on curved spacetimes and 
 spin-base transformations for left-handed and right-handed Weyl spinors decouple.

The covariant derivative $\nabla_\mu$ in Eq.~\eqref{eq:fermion_kinetic_action} contains the spin connection for Weyl fermions but lacks the gauge fields as we only consider global charges.
For $\psi$ left-handed and $\chi$ right-handed it reads
\begin{align}
    \nabla_\mu \psi = \left(\partial_\mu + \frac{1}{8} \omega_\mu{}^{\rho\sigma} (\sigma_\rho \barsigma_\sigma - \sigma_\sigma \barsigma_\rho)\right) \psi\,, \nn\\
    \nabla_\mu \chi = \left(\partial_\mu + \frac{1}{8} \omega_\mu{}^{\rho\sigma} (\barsigma_\rho \sigma_\sigma - \barsigma_\sigma \sigma_\rho)\right) \chi\,.
\end{align} 
Concerning the metric perturbation in terms of gravitons, this entails
\begin{align}
    \barsigma^\mu[g] = \barsigma^\mu[\bar{g}] - \frac{1}{2} \barsigma^\nu[\bar{g}] h^\mu{}_\nu + \frac{3}{8} \barsigma^\rho[\bar{g}] h_{\rho\sigma} h^{\mu\sigma} + \mathcal{O}(h^3)\,, \nn\\
    \sigma^\mu[g] = \sigma^\mu[\bar{g}] - \frac{1}{2} \sigma^\nu[\bar{g}] h^\mu{}_\nu + \frac{3}{8} \sigma^\rho[\bar{g}] h_{\rho\sigma} h^{\mu\sigma} + \mathcal{O}(h^3)\,,
\end{align}
where the gamma matrices on the right hand side are evaluated on the background.

\section{General setup}
\label{app:general_setup}

Throughout this paper, we work in Euclidean signature.

The full truncation of Eq.~\eqref{eq:total_action} is composed of the four-fermion vertices and the fermionic terms described in \eqref{eq:4_fermion_action} and \eqref{eq:fermion_kinetic_action} in combination with the gravitational action, gauge fixing and Faddeev--Popov ghosts.
For the gravitational sector we employ the Einstein--Hilbert action given in \eqref{eq:EH_action}.
Diffeomorphism symmetry is fixed by
\begin{equation}
\label{eq:gauge_fixing_action_euclidean}
	\Gamma_k^{\text{g.f.}} = \frac{1}{2} \int\dd[4]{x \sqrt{\bar{g}}} \mathcal{F}_{\mu} \bar{g}^{\mu\nu} \mathcal{F}_{\nu}\,,
\end{equation}
with the gauge fixing condition 
\begin{equation}
	\label{eq:gauge_fixing_condition}
	\mathcal{F}_{\mu} = \mathcal{F}_\mu{}^{\rho\sigma}h_{\rho\sigma}
	= \frac{1}{\sqrt{16\pi G \alpha}} \left[\delta_{\phantom{(}\mu}^{(\rho}\bar{\nabla}^{\sigma)} - \frac{1+\beta}{4}\bar{g}^{\rho\sigma} \bar{\nabla}_\mu\right] h_{\rho\sigma}\,.
\end{equation}
For completeness, we note that the Faddeev--Popov ghosts term reads
\begin{equation}
\label{eq:ghost_action_euclidean}
	\Gamma_k^{\text{gh.}} = \int\dd[4]{x \sqrt{\bar{g}}} \bar{c}^\mu \mathcal{F}_{\mu}{}^{\rho\sigma} (\nabla_{\rho} c_{\sigma} + \nabla_{\sigma} c_{\rho})\,.
\end{equation}
There is no direct coupling between fermions and ghosts in our truncation, although such a coupling may  be induced \cite{Eichhorn:2013ug}, raising interesting interpretational questions for the status of gauge fixing.
 We take Landau--DeWitt gauge $\beta \to \alpha \to 0$, which is a fixed point of the renormalization group flow \cite{Knorr:2017fus}.

We choose the regulators according to
\begin{equation}
	\label{eq:regulator_tensor}
	R_{k,AB}(p^2) = \left(\Gamma_{k,AB}^{(2)}(p^2) - \Gamma_{k,AB}^{(2)}(0)\right)\bigg|_{\Phi=0} r_{AB}(p^2/k^2)\,,
\end{equation}
with shape functions $r_{AB}(p^2/k^2)$.
We pick  the Litim  shape functions \cite{Litim:2001up}
\begin{align}
	\label{eq:shape_function}
	r_F(x=p^2/k^2) &= \left(\frac{1}{\sqrt{x}} - 1\right) \Theta(1-x)\,, \nn\\
	r_h(x=p^2/k^2) &= \left(\frac{1}{x} - 1\right) \Theta(1-x)\,.
\end{align}
for the fermion and the graviton respectively.

The main calculations were performed in \texttt{Wolfram Mathematica}, using the package suite \texttt{xAct} \cite{Martin-Garcia:2008ysv, Brizuela:2008ra, xAct_webpage} for tensor algebra and employing \texttt{FieldsX} \cite{Frob:2020gdh} to deal with spinors.

\section{Fixed points of the fermionic subsystem}
\label{app:fixed_points}

\setlength{\tabcolsep}{5pt}
\begin{table}[h!]
	\footnotesize 
	\begin{center}
		\begin{tabular}{ l | r r r r r | r r r r r r r }
        No. & $c_{qq\,*}$ & $c_{qu\,*}$ & $c_{qd\,*}$ & $c_{ud\,*}$ & $c_{quqd\,*}^+$ & $\theta_1$ & $\theta_2$ & $\theta_3$ & $\theta_4$ & $\theta_5$ & $\theta_6$ & $\theta_7$ \\ \hline \hline
			1 & $0$ & $0$ & $0$ & $0$ & $0$ & $-2.0$ & $-2.0$ & $-2.0$ & $-2.0$ & $-2.0$ & $-2.0$ & $-2.0$ \\ \hline
			2 & $-158.$ & $0$ & $0$ & $0$ & $0$ & $4.0$ & $4.0$ & $2.0$ & $-2.0$ & $-2.0$ & $-2.0$ & $-2.0$ \\ \hline
			3 & $-79.0$ & $-2c_{qq\,*}$ & $0$ & $0$ & $0$ & $2.0$ & $2.0$ & $-1.0$ & $-1.0$ & $-2.0$ & $-6.0$ & $-8.0$ \\
			4 & $-79.0$ & $0$ & $-2c_{qq\,*}$ & $0$ & $0$ & $2.0$ & $2.0$ & $-1.0$ & $-1.0$ & $-2.0$ & $-6.0$ & $-8.0$ \\ \hline
			5 & $-7.97$ & $-69.1$ & $0$ & $0$ & $0$ & $2.0$ & $0.63$ & $-1.0$ & $-1.9$ & $-2.0$ & $-2.3$ & $-2.6$ \\
			6 & $-7.97$ & $0$ & $-69.1$ & $0$ & $0$ & $2.0$ & $0.63$ & $-1.0$ & $-1.9$ & $-2.0$ & $-2.3$ & $-2.6$ \\ \hline
      7 & $0$ & $0$ & $0$ & $-316.$ & $0$ & $4.0$ & $2.0$ & $0$ & $-2.0$ & $-2.0$ & $-2.0$ & $-10.$ \\ \hline
			8 & $0$ & $0$ & $0$ & $79.0$ & $0$ & $2.0$ & $-1.2$ & $-2.0$ & $-2.0$ & $-2.0$ & $-2.3$ & $-2.5$ \\ \hline
      9 & $-158.$ & $0$ & $0$ & $2c_{qq\,*}$ & $0$ & $10.$ & $6.0$ & $2.0$ & $2.0$ & $-2.0$ & $-2.0$ & $-10.$ \\ \hline
			10 & $-158.$ & $0$ & $0$ & $79.0$ & $0$ & $4.8$ & $3.8$ & $2.0$ & $2.0$ & $-2.0$ & $-2.0$ & $-2.5$ \\ \hline
      11 & $-88.7$ & $111.$ & $111.$ & $79.0$ & $0$ & $4.0$ & $2.0$ & $0.48$ & $-1.3$ & $-3.6$ & $-5.1$ & $-10.$ \\ \hline
      12 & $-52.6$ & $-2c_{qq\,*}$ & $-2c_{qq\,*}$ & $2c_{qq\,*}$ & $0$ & $2.0$ & $-0.67$ & $-0.67$ & $-4.7$ & $-4.7$ & $-4.7$ & $-10.$ \\ \hline
			13 & $-16.0$ & $-67.3$ & $-67.3$ & $79.0$ & $0$ & $3.1$ & $2.9$ & $2.6$ & $2.0$ & $-1.9$ & $-2.3$ & $-3.1$ \\ \hline
			14 & $-13.5$ & $75.5$ & $-45.5$ & $-300.$ & $0$ & $4.9$ & $2.0$ & $0.17$ & $-2.0$ & $-2.9$ & $-3.1$ & $-9.6$ \\
			15 & $-13.5$ & $-45.5$ & $75.5$ & $-300.$ & $0$ & $4.9$ & $2.0$ & $0.17$ & $-2.0$ & $-2.9$ & $-3.1$ & $-9.6$ \\ \hline
			16 & $-11.7$ & $-58.5$ & $-58.5$ & $2c_{qq\,*}$ & $0$ & $2.4$ & $2.0$ & $1.2$ & $-1.7$ & $-2.3$ & $-2.6$ & $-2.6$ \\ \hline
			17 & $-6.76$ & $-2c_{qq\,*}$ & $-c_{ud\,*}$ & $62.5$ & $0$ & $2.0$ & $1.2$ & $-0.14$ & $-1.8$ & $-2.2$ & $-2.3$ & $-2.7$ \\ \hline
			18 & $-6.76$ & $-c_{ud\,*}$ & $-2c_{qq\,*}$ & $62.5$ & $0$ & $2.0$ & $1.2$ & $-0.14$ & $-1.8$ & $-2.2$ & $-2.3$ & $-2.7$ \\ \hline
			19,\,20 & $-2.22$ & $-26.3$ & $-26.3$ & $-4.45$ & $\pm49.6$ & $2.0$ & $-0.71$ & $-1.7$ & $-1.9$ & $-2.1$ & $-2.2$ & $-3.1$ \\ \hline
			21,\,22 & $-2.22$ & $-26.3$ & $-26.3$ & $79.0$ & $\pm54.5$ & $2.2$ & $2.0$ & $0.0011$ & $-1.8$ & $-2.1$ & $-2.4$ & $-3.5$ \\ 
		\end{tabular}
    \caption[List of partially decoupled vectorial-axial fixed points]{List of fixed points (FPs) for the purely fermionic system.
		The fixed points are numbered arbitrarily.
			We provide fixed point values, with critical exponents $\theta_I$ sorted in decreasing order.
            The fixed point values $c_{uu\,*} = -v_4(c_{qu\,*}^2+c_{ud\,*}^2/2)$ and $c_{dd\,*} = -v_4(c_{qd\,*}^2 + c_{ud\,*}^2/2)$ are omitted.
		We omit the horizontal line between the pairs $(3,4)$, $(5,6)$ and $(14,15)$ because they are symmetric under the swap $u \leftrightarrow d$. }
		\label{tab:fermionic_fps}
	\end{center}
\end{table}

The structure between coupling values visible in some  of the fixed points can be understood by considering the different subsystems the truncation Eq.~\eqref{eq:4_fermion_action} features.
Removing a field completely from the interactions means some couplings must be zero but the remaining two fields generate a viable system with inherent fixed points.
For example, removing $d$ as a dynamical field, all interactions containing $d$ must be deactivated, i.e.\ $\cqd = \cud = \cdd = \cRe = 0$ or at least decoupled (in this case, $\cdd$ is not necessarily zero but $d$ does not  interact with $u$ or $q$), fixed points 2 and 3 are of the latter type.
Moreover, one can project on subsystems that respect a higher degree of symmetry than $\mathrm{SU}(2)_L\times \mathrm{U}(1)_Y$.
One can identify a ladder of cascading symmetry constraints up to the maximal symmetry of the kinetic term discussed in secions~\ref{sec:action} and \ref{sec:categories}.
In particular, we find
\begin{equation}
\begin{aligned}
    \label{eq:symmetry_ladder}
    &\mathrm{SU}(2)_L\times\mathrm{U}(1)_Y \\
    &\quad\downarrow\; c_{uu} = c_{dd}, c_{qu}=c_{qd} \\
    &\mathrm{SU}(2)_L\times \mathrm{S}_2^{ud}\times \mathrm{U}(1)_Y \quad\text{(realized by FPs 8,10,11,13,21,22)} \\
    &\quad\downarrow\; {c_{ud} = 2c_{uu}} \\
    &\mathrm{SU}(2)_L\times\mathrm{SU}(2)_R\times \mathrm{U}(1)_Y \quad\text{(FPs 19,20)} \\
    &\quad\downarrow\; {\cRe=0} \\
    &\mathrm{SU}(2)_L \times \mathrm{SU}(2)_R \times \mathrm{U}(1)^4 \quad\text{(FPs 2,7)}\\
    &\quad\downarrow\; {c_{qq} = c_{uu}} \\
    & \mathrm{SU}(4)\times \mathrm{U}(1)^4 \quad\text{(FPs 9, 16)} \\
    &\quad\downarrow\; {c_{qu}=-c_{ud}} \\
        &\mathrm{U(4)} \quad\text{(FPs 1,12)}\,.
\end{aligned}
\end{equation}
Reading the list from top to bottom, we project on a subspace with increasingly larger symmetry.
 Conversely, reading from bottom to top, these symmetry groups are broken  to subgroups or lifted  entirely.
Each fixed point of a higher symmetry group naturally fulfills the conditions of the weakened constraint above.
$S_2^{ud}$ is the swap of $u$ and $d$.
$\mathrm{SU}(2)_R$ denotes a special unitary transformation of the doublet $(u,d)$.
$\mathrm{U}(1)^4$ is the individual phase rotation of each field component, generalizing the joint rotation with a fixed ratio of charges defined by $\mathrm{U}(1)_Y$.
The overview \eqref{eq:symmetry_ladder} could be extended to show more symmetries that have $\mathrm{SU}(2)_L\times\mathrm{U}(1)_Y$ as a subgroup and populated by the corresponding fixed points of Tab.~\ref{tab:fermionic_fps}.

The extensions of the fixed points 21 and 22 are the ones that collide with the SGFP at $g_{*\,\text{crit.}}(\lambda)$,  see Fig.~\ref{fig:perturbativity}, giving the SGFP a single relevant direction towards $\cRe$.

\section{Extended Plots}
\label{app:extended_plots}

\begin{figure}[!ht]
\begin{center}
    \includegraphics[width=0.75\textwidth]{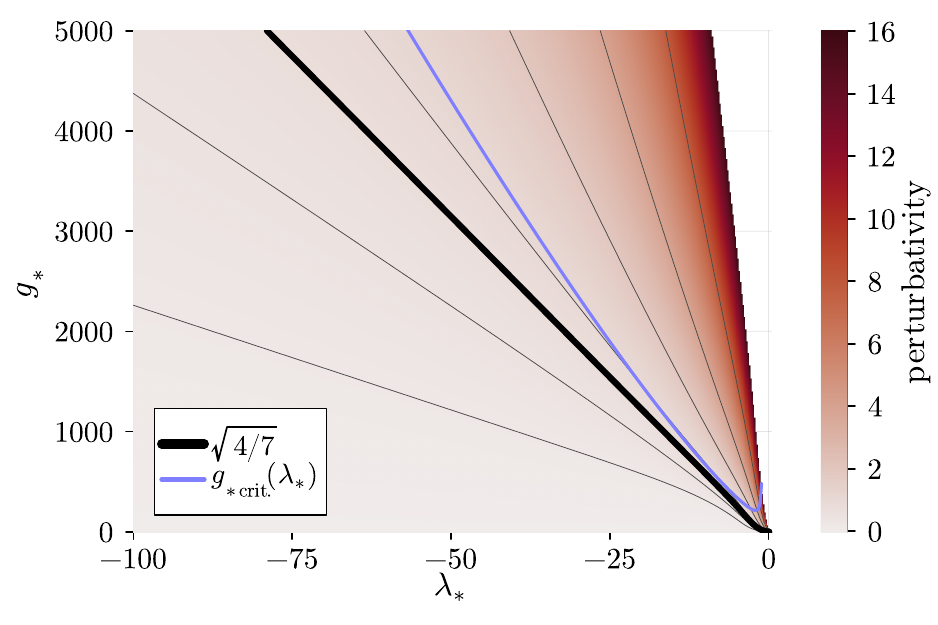}
    \caption{Perturbativity, defined in Eq.~\eqref{eq:def_perturbativity} of the shifted Gaussian fixed point. 
If making the coupling $\cRe$ relevant was the only effect of anomalous quantum scaling, then the minimum perturbativity would be $\sqrt{4/7}$, which is marked alongside the collision boundary.}
\label{fig:perturbativity_extended}
\end{center}
\end{figure}

\begin{figure}[!ht]
\begin{center}
    \includegraphics[width=0.75\textwidth]{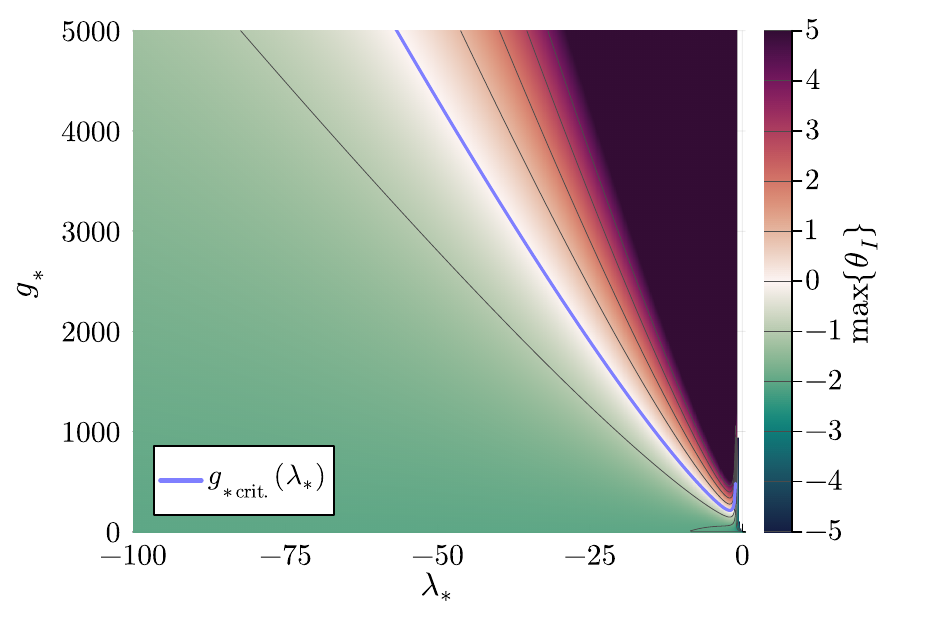}
    \caption{Maximum critical exponent of the shifted Gaussian fixed point for extended gravitational parameters.}
\label{fig:max_theta_extended}
\end{center}
\end{figure}

In the analysis of the shifted Gaussian fixed point (see Sec.~\ref{sec:FP_collisions}), the shown collision boundary (Figs.~\ref{fig:perturbativity}, \ref{fig:perturbativity_extended} and \ref{fig:max_theta_extended}) of the SGFP (for $\lambda > -83.581$) is given by 
\begin{equation}
	\label{eq:collision_boundary}
    \begin{aligned}
        g_{*\,\text{\text{crit}.}}(\lambda) = \frac{20}{A} \bigg[&5 \sqrt{3} \pi \sqrt{- \left(3 -4 \lambda \right)^8 (2 \lambda -1)^5} \\
					      &- 2 \pi \big(3 + 2 \lambda  \left(-5 + 4 \lambda \right)\big)^2 \Big(87 + \lambda  \big(-195 + 2 \lambda (43 + 6 \lambda )\!\big)\!\Big)\bigg]\,,
    \end{aligned}
\end{equation}
with
\begin{equation}
    A = \Bigg(24201 + 2 \lambda  \bigg(-45585 + 2 \lambda  \Big(20589 + 4 \lambda  \big(3837 + \lambda  \left(-7721 + 36 \lambda  \left(81 + \lambda \right)\right)\!\big)\!\Big)\!\bigg)\!\Bigg)\,.
\end{equation}
\end{document}